# Development of the Stochastic Interpretation of Quantum Mechanics by E. Nelson. Derivation of the Schrödinger-Euler-Poisson Equations

## Mikhail Batanov-Gaukhman

Ph.D., Associate Professor, Institute No. 2 "Aircraft, rocket engines and power plants",
Federal State Budgetary Educational Institution of Higher Education "Moscow Aviation Institute (National Research University)", Volokolamsk highway 4, Moscow, Russian Federation (alsignat@yandex.ru)

**Abstract:** The aim of the article is to develop the stochastic interpretation of quantum mechanics by E. Nelson on the basis of balancing the intra-systemic contradiction (i.e., antisymmetry) between "order" and "chaos". For the set task, it is proposed to combine two mutually opposite system-forming principles: "the principle of least action" and "the principle of maximum entropy" into one the "principle of averaged efficiency extremum". In a detailed consideration of the averaged states of a chaotically wandering particle, the time-independent (stationary) and time-dependent stochastic Schrödinger-Euler-Poisson equations are obtained as conditions for finding the extremals of the globally averaged efficiency functional of the stochastic system under study. The resulting stochastic equations coincides with the corresponding Schrödinger equations up to coefficients. In this case, the ratio of the reduced Planck constant to the particle mass is expressed through the averaged characteristics of a three-dimensional random process in which the considered wandering particle participates. The obtained stochastic equations are suitable for describing the quantum states of stochastic systems of any scale.

**List of abbreviations and definitions**
MSQM is the mass-independent stochastic quantum mechanics;
QM is the quantum mechanics;
PSRP is the pseudo-stationary random process;
SQM is the stochastic quantum mechanics;
SRP is the stationary random process;
PA is the probability amplitude;
PDF is the probability density function;
ChWP is the chaotically wandering particle;
Pico-particle is a particle or antiparticle with a size of ~ $10^{-8} – 10^{-13}$ cm;
Micro-particle is a particle with dimensions of ~ $10^{-7} – 10^{-3}$ cm;
Macro-particle is a compact bodies with dimensions of ~ $10^{-2} – 10^{4}$ cm;
$s = S/m$ is the "efficiency" of the particle with mass $m$;
$\varepsilon = E/m$ is the "mechanical energiality" of the particle with mass $m$;
$u = U/m$ is the "potential energiality" of the particle with mass $m$;
$k = T/m$ is the "kinetic energiality" of the particle with mass $m$.

## 1 BACKGROUND AND INTRODUCTION

In modern physics, there are several dozen interpretations of Quantum Mechanics (QM). Each of them has its own advantages and disadvantages, but none of them is precisely defined, since many researchers often put different meanings into the same concepts.

One of the reasons for this situation in quantum physics is associated with a different attitude to the wave function $\Psi(x,t)$.

Most experts agree with M. Born's statement that the square of the modulus of the wave function of a particle $\Psi(x,t)$ is equal to the probability density function (PDF) of the particle's location at a point $x$

$$|\Psi(x,t)|^2 = \rho(x,t).$$

However, it should be borne in mind that, in general, this PDF is a complex function of several factors associated with the measurement process.

$$|\Psi(x,t)|^2 = \rho(x,t) = f[\rho_p(x,t), \rho_m(x,t), \rho_e(x,t), \rho_d(x,t), \rho_o(x,t)], \qquad (0.1)$$

where

$\rho_p(x,t)$ is the PDF associated with the chaotic behavior of the particle;

$\rho_m(x,t)$ is the PDF, associated with the method errors;

$\rho_e(x,t)$ is the PDF, associated with the influence of the external environment;

$\rho_d(x,t)$ is the PDF, associated with the instrument errors;

$\rho_o(x,t)$ is the PDF, associated with the operator errors.

An example of functional dependence (0.1) is the PDF

$$|\Psi(x,t)|^2 = \rho(x,t) = \frac{1}{\sqrt{2\pi[\sigma_{px}^2 + \sigma_{mx}^2 + \sigma_{ex}^2(t) + \sigma_{dx}^2 + \sigma_{ox}^2]}} \exp\left\{-\frac{x^2}{2[\sigma_{px}^2 + \sigma_{mx}^2 + \sigma_{ex}^2(t) + \sigma_{dx}^2 + \sigma_{ox}^2]}\right\},$$

where (0.2)

$$\sigma_{ix}^2 = \int_{-\infty}^{+\infty} \rho_i(x) x^2 dx \quad (\text{here } i = p, m, d, o), \qquad \sigma_{ex}^2(t) = \int_{-\infty}^{+\infty} \rho_e(x,t) x^2 dx$$

is a variance of the $i$-th influencing factor on the measurement result.

All of the above factors are present when measuring the physical characteristics of particles of any scale. However, depending on the particle size, these factors affect the result differently.

At the same time, almost all specialists who study the properties of non-relativistic pico-particles (i.e., particles with characteristic sizes of atomic and subatomic scales, $10^{-8} - 10^{-13}$ cm) use the same mathematical apparatus of quantum mechanics (QM), designed to predict possible configurations and evolutions of wave functions $\Psi(x,t)$ of one particle or an ensemble of identical particles.

Focusing on certain factors influencing the measurement process, using the same mathematical apparatus, leads to the development of different interpretations of QM.

For example, in a number of experiments pico-particles are extremely sensitive to the influence of the measuring system and the observer on them. The effects associated with the fact that the operator, by his presence, introduces parasitic capacitance and inductance into radio-electronic devices, which leads to an imbalance of electronic resonators, etc., are widely known.

In this case, the wave function (0.1) should take into account all influencing factors, while the methodology for perceiving the results obtained is most consistent with the Copenhagen interpretation of QM, formulated by Niels Bohr and Werner Heisenberg.

In other experiments, the factors that interfere with the measurement are so insignificant that they can be neglected. For example, we judge the possible states of an electron in a hydrogen atom by its emission spectrum. If we abstract from the slight broadening of the spectral lines associated with the influence of various accompanying factors, then in this case PDF (0.1) takes the form

$$|\Psi(x,t)|^2 = \rho(x,t) = f[\rho_p(x,t)\, \rho_e(x,t)].$$

This wave function characterizes only the properties of the electron itself, taking into account the influence of the vacuum, leading to the Lamb shift of the spectral lines.

In this article, we will adhere to the Stochastic interpretation of quantum mechanics, most clearly expressed in the works of Edward Nelson [2,3,4], published in 1966 – 1985.

In addition to E. Nelson, this interpretation of QM was developed by R. Fürth [5], I. Fényes [6], W. Weizel [7], M. Pavon, [8], K. Namsrai [28]. An alternative stochastic interpretation of QM was developed by R. Tsekov [9].

Among the more recent works on stochastic quantum mechanics, the articles by J. Lindgren & J. Liukkonen (2019) [29] and T. Koide & T. Kodama (2018) [30] should be noted.

Nelson's stochastic interpretation is associated with the logical construction of QM by analogy with the theory of Brownian motion [more precisely, the Ornstein-Uhlenbeck process].

In Nelson's interpretation, the reason for the chaotic behavior of a pico-particle is associated with the effect of vacuum fluctuations on it. The diffusion coefficient of such a stochastic process turns out to be imaginary due to the absence of friction and the specificity of the vacuum viscosity. Thus, in the stochastic interpretation of QM, the primary is not the wave function $\Psi(x,t)$, but complex small-scale curvatures of the space-time continuum (i.e., the Wheeler-Bohm-Vigier "quantum foam"), which affect to the colloidal pico-particle.

In this case, PDF (0.1) takes the simplest form

$$|\Psi(x,t)|^2 = \rho(x,t) = \rho_e(x,t) = \psi(x,t)\,\psi^*(x,t), \qquad (0.3)$$

since it characterizes only the chaotic behavior of a structureless particle under the influence of a turbulent environment.

Recall that the Langevin and Fokker-Planck stochastic equations describe Brownian motion without taking into account the structure of particles and the uncertainty associated with measurement errors. However, there is a fundamental difference between Brownian particles (~ $10^{-4}$ cm in size) and pico-particles (~ $10^{-8} - 10^{-13}$ cm in size). Brownian (colloidal) particles can be observed with a microscope, practically without affecting them, while pico-particles, in principle, cannot be observed directly.

In this article, the maximally simplified (more precisely, not taking into account the measurement error and the influence of other particles) probability amplitude (PA) $\Psi(x,t) = \psi(x,t)$ will be called the "pure" wave function.

It should be noted that in most books and textbooks on quantum mechanics, initially the PA $\Psi(x,t)$ means the "pure" wave function of the particle. This is one of the reasons for the lack of agreement between theorists and experimenters, as well as between specialists working in various fields of quantum physics. Apparently, it was the attitude to the "pure" or "impure" wave function $\Psi(x,t)$ that caused the disputes between A. Einstein (who studied Brownian motion in his youth) and N. Bohr (whose early works were associated with the atomic emission spectra).

So, in this article, under the stochastic interpretation of quantum mechanics by Edward Nelson, we mean a version of QM in which the wave function $\psi(x,t)$ characterizes only the chaotic behavior of a wandering particle under the influence of environmental fluctuations, without taking into account measurement errors and the influence of the operator. This particle (like a Brownian corpuscle) has a volume and a chaotic trajectory of motion. In this case, the wave function $\psi(x,t)$ has the statistical character attributed to it by M. Born.

At the same time, it is taken into account that within the framework of the Nelson's stochastic interpretation of the QM, the "pure" PA $\psi(x,t)$ turns out to be a kind of "intellectual thing-in-itself". This is because the "pure" wave function $\psi(x,t)$ can be found out only by solving stochastic differential equations. Any attempt to perform a measurement will lead to a partial distortion or complete destruction of the stochastic system under study, and hence to a change in its PA $\psi(x,t)$.

This article attempts to develop the foundations of mass-independent stochastic quantum mechanics (MSQM), which is a development of Nelson's stochastic quantum mechanics (SQM) [2,3], and proposes a solution to the problem of measuring "pure" parameters of stochastic quantum systems.

The article attempts to exclude mass from all the physical quantities and constants mentioned here. This is one of the steps in the program of complete clearing of physics from the concept of "mass" and its dimension "kilogram" (or "pound", etc.), since this phenomenological concept, in the author's opinion, is superfluous and hinders the development of fully geometrized physics.

A probabilistic model of a chaotically wandering particle (ChWP) is considered, which, like the pico-particle by E. Nelson [2], has a volume and a continuous trajectory of motion. But in contrast to the SQM [2], in the MSQM there are no restrictions on the size of the investigated particle. Based on this model (by the calculus of variations method, different from [2]), stochastic Eq.s (49), (52), (77), (78a) are derived, which are a generalization of the Schrödinger equations, and also was obtained Eq. (83) corresponding to the form of the diffusion equation (a particular case of the Fokker-Planck-Kolmogorov equation).

In the author's opinion, the main advantages of the obtained stochastic equations are that they are suitable for describing the averaged quantum states of chaotically wandering particles (ChWP) of any scale and quality. For example, this approach is suitable for describing quantum effects when averaging the chaotic displacements of the geometric (or hypothetical) center: the tip of a tree branch under the influence of wind blows, or random displacements of the center of the planet's core, or changes in stock prices on the stock exchange, or displacement concentration of the student's attention from the subject of study, etc.

## 2 METHOD

Stochastic objects (systems) are formed while simultaneous striving for "order" and "chaos" [16, 26]. This is due to the influence on any stochastic object of two mutually opposite system-forming factors: the "principle of least action" and the "principle of maximum entropy". The mutually opposite tendency of stochastic systems to "order" and "chaos" leads them to equilibrium, optimization, energy saving, energy efficiency and space-time symmetry.

In this article, an attempt is made to combine the two above-mentioned mutually opposite (antisymmetric) principles into one the "principle of averaged efficiency extremum" (PAEEx), which is internally balanced with respect to "determinism" (i.e., predetermination) and "randomness". This new principle, as a result, theoretically substantiates the empirically manifested advantage of symmetric states of the free stochastic objects.

### 2.1 Probabilistic model of a wandering particle

Consider a solid particle with a volume and a trajectory of motion (Figure 1). Let's assume that under the influence of fluctuations in the environment, the particle constantly wanders chaotically in the vicinity of the conditional center, combined with the origin of the fixed coordinate system *XYZ*.

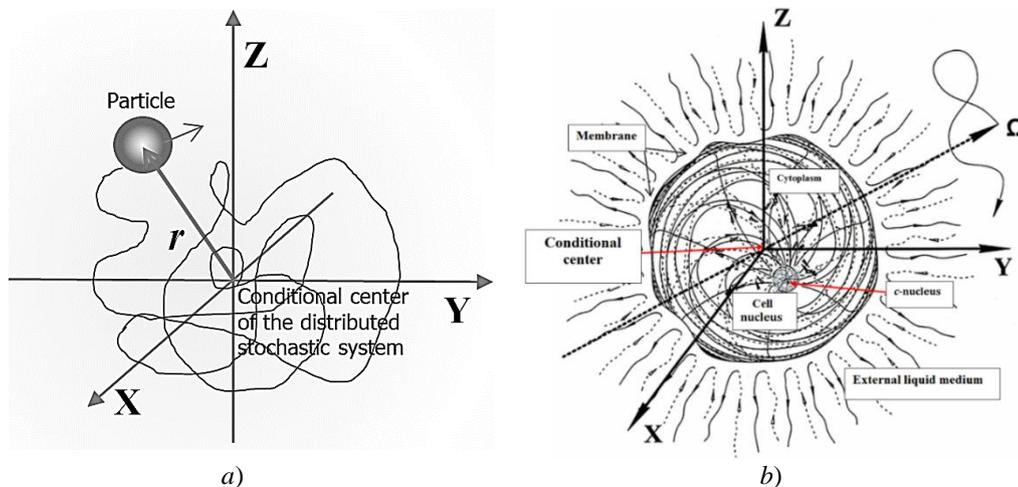

*a)* *b)*

**Fig. 1.** *a*) A simplified model of a solid particle (more precisely, a geometric or hypothetical center of a compact object), which randomly wanders around a conditional center due to random force effects from a fluctuating environment. *b*) Examples of a chaotically wandering particle (ChWP), when deviating from the center, stresses arise in the environment that seek to return the particle to the central position

Random displacements in three spatial (or conditional, phase) measurements with time of **the geometric (or hypothetical) center** of the following objects or subjects can be classified as examples of a chaotically wandering particle (ChWP): a valence electron in a hydrogen-like atom; an oscillating atom in a crystal lattice; a fluttering heart in the chest of an animal; a fluttering yolk in a chicken egg; a fluttering moth in the vicinity of a burning lamp; a fish floating in the aquarium; a shifting mosquito swarm; a trembling organelle in the cytoplasm; a vibrating incandescent core in the bowels of the planet; a wandering pollen in diluted sugar syrup; a wiggling embryo in the womb; a shifting school of fish in the ocean; a fluttering flower in the wind; a fluctuating cost of service in the labor market; changes in public opinion regarding a social problem; distractions from a problem which a scientist is trying to solve; etc.

The random behavior of most of 3D objects and subjects around us, in the presence of a clearly expressed geometric (or hypothetical) center and a sufficiently long observation period, can be interpreted as the behavior of a chaotically wandering particle (ChWP) in a 3D real (or phase) space. Therefore, the conclusions made in this article are of a universal nature and apply to stochastic objects and subjects of any scale and quality.

Let's consider the ChWP as a stochastic system distributed in a 3-dimensional space, with a selected center at the origin of the coordinate system (see Figure 1).

We assume that the total mechanical energy of the ChWP $W(v,x,y,z,t)$ at the point with coordinates $x,y,z$ and at time $t$ is equal to

$$W(v,x,y,z,t) = T(v,x,y,z,t) + U(x,y,z,t) \pm \varepsilon(x,y,z,t), \qquad (1)$$

where $T(v,x,y,z,t)$ is the kinetic energy of a particle, which is determined by the velocity $v(v_x, v_x, v_x, x, y, z, t)$, at the time $t$ at the point with coordinates $x,y,z$ (Figure 1);

$U(x,y,z,t)$ is the potential energy of a particle associated with the elasticity of its environment (or other factors) that, on average, tend to return this particle to the conditional center of the considered distributed stochastic system;

$\varepsilon(x,y,z,t)$ is the energy that a particle receives or gives away for a short time when interacting with a chaotically fluctuating environment. This interaction energy can initially be spent on elastic deformations and/or on a change in the angular velocity of rotation of a particle, but then it, in different fractions, randomly transforms into kinetic energy $T(v,x,y,z,t)$ and/or into potential energy $U(x,y,z,t)$ of the ChWP.

We represent expression (1) as

$$W(v,x,y,z,t) \mp \varepsilon(x,y,z,t) = T(v,x,y,z,t) + U(x,y,z,t), \qquad (2)$$

Let's denote the left side of this expression as follows

$$E(x,y,z,t) = W(v,x,y,z,t) \mp \varepsilon(x,y,z,t) \qquad (3)$$

and we will call it the "mechanical energy" of the ChWP.

In this case, all three energies of the ChWP: $E(x,y,z,t)$, $T(v,x,y,z,t)$ and $U(x,y,z,t)$ are random variables, but such that in each point of the region of space under consideration satisfies the averaged equality

$$<E(x,y,z,t)> = <T(v,x,y,z,t)> + <U(x,y,z,t)>, \qquad (4)$$

or  $\quad <T(v,x,y,z,t)> + <U(x,y,z,t)> - <E(x,y,z,t)> = 0, \qquad (5)$

where $<>$ means local averaging, i.e. averaging in the neighborhood of a point with coordinates $x,y,z$, located in the interior of a distributed stochastic system (see Figure 1).

If the speed of movement of the ChWP $v(v_x,v_y,v_z)$ is small comparing to the speed of light, then according to non-relativistic mechanics, it has a kinetic energy

$$T\left(p_x, p_y, p_z, x, y, z, t\right) = \frac{p_x^2(x,y,z,t) + p_y^2(x,y,z,t) + p_z^2(x,y,z,t)}{2m}, \qquad (6)$$

where $p_i = mv_i(x,y,z,t)$ are the instantaneous value of the $i$-th component of the ChWP momentum vector at time $t$ and at a point with coordinates $x,y,z$ (Figure 1); $m$ is the mass of the particle.

Let's integrate Eq. (5) with respect to time

$$\int_{t_1}^{t_2} \left[ <T(p_x, p_y, p_z, x, y, z, t)> + <U(x, y, z, t)> - <E(x, y, z, t)> \right] dt = 0. \qquad (7)$$

and we will call this expression "locally averaged balance of a stochastic system".

Let's write expression (7) in a compact form

$$\int_{t_1}^{t_2}[<T(\vec{p},\vec{r},t)> +<U(\vec{r},t)> -<E(\vec{r},t)>]\,dt = 0, \qquad (8)$$

where  $\qquad \vec{p}:=(p_x,p_y,p_z), \qquad \vec{r}:=(x,y,z),$

we also introduce the notation  $\hfill (9)$

$$|\vec{r}\,|^2 := x^2+y^2+z^2, \quad d\vec{r}:=dxdydz, \quad \int_{-\infty}^{\infty} d\vec{r} := \int_{-\infty}^{\infty}\int_{-\infty}^{\infty}\int_{-\infty}^{\infty} dxdydz,$$

$$|\vec{p}\,|^2 := p_x^2+p_y^2+p_z^2, \quad d\vec{p}:=dp_xdp_ydp_z, \quad \int_{-\infty}^{\infty} d\vec{p} := \int_{-\infty}^{\infty}\int_{-\infty}^{\infty}\int_{-\infty}^{\infty} dp_xdp_ydp_z.$$

**2.2 Globally averaged efficiency of the ChWP**

Let's perform a global averaging of the locally averaged balance (8) over the entire area, in which the ChWP moves (see Figure 1)

$$\overline{<S_r(t)>} = \overline{\int_{t_1}^{t_2}[<T(\vec{p},\vec{r},t)> +<U(\vec{r},t)> -<E(\vec{r},t)>]\,dt}. \qquad (10)$$

The integration and averaging operations are commutative, so expression (10) can be represented as

$$\overline{<S_r(t)>} = \int_{t_1}^{t_2}[\overline{<T(\vec{p},\vec{r},t)>} +\overline{<U(\vec{r},t)>} -\overline{<E(\vec{r},t)>}]dt. \qquad (11)$$

We represent the globally averaged kinetic energy of the ChWP in the following form

$$\overline{<T(\vec{p},\vec{r},t)>} = \frac{1}{2m}\int_{-\infty}^{\infty}\int_{-\infty}^{\infty}\int_{-\infty}^{\infty}\rho(p_x,p_y,p_z,t)[p_x^2+p_y^2+p_z^2]dp_xdp_ydp_z, \quad (12)$$

where $\rho(p_x,p_y,p_z,t)$ is a joint probability density function (PDF) for 3 components of the particle momentum vector $p_x, p_y, p_z$. In general, $\rho(p_x,p_y,p_z,t)$ may change over time *t*.

Let us represent the globally averaged potential energy $\overline{<U(\vec{r},t)>}$ and the globally averaged mechanical energy $\overline{<E(\vec{r},t)>}$ in the following form

$$\overline{<U(\vec{r},t)>} = \int_{-\infty}^{\infty}\int_{-\infty}^{\infty}\int_{-\infty}^{\infty}\rho(x,y,z,t)<U(x,y,z,t)>dxdydz, \qquad (13)$$

$$\overline{<E(\vec{r},t)>} = \int_{-\infty}^{\infty}\int_{-\infty}^{\infty}\int_{-\infty}^{\infty}\rho(x,y,z,t)<E(x,y,z,t)>dxdydz, \qquad (14)$$

where $\rho(x,y,z,t)$ is the joint PDF of the projections of the ChWP position *x,y,z* on the *X,Y,Z* axes (see Figure 1). In general, $\rho(x,y,z,t)$ may change over time *t*.

Substituting globally averaged values (12), (13) and (14) into Ex. (11), taking into an account the notation (9), we obtain a globally averaged balance of the ChWP

$$\overline{<S_r(t)>} = \int_{t_1}^{t_2} \left\{ \frac{1}{2m} \int_{-\infty}^{\infty} \rho(\vec{p},t)|\vec{p}|^2 d\vec{p} + \int_{-\infty}^{\infty} \rho(\vec{r},t)[<U(\vec{r},t)> - <E(\vec{r},t)>]d\vec{r} \right\} dt. \quad (15)$$

Let's exclude the superfluous concept of "mass" from the proposed mathematical model. To do this, we introduce quantities that do not depend on the mass:

$$\overline{<s_r(t)>} = \frac{\overline{<S_r(t)>}}{m} \quad (16)$$

is a globally averaged "efficiency" of the ChWP;

$$<\varepsilon(\vec{r},t)> = \frac{<E(\vec{r},t)>}{m} \quad (17)$$

is a locally averaged "mechanical energiality" of the ChWP;

$$<u(\vec{r},t)> = \frac{<U(\vec{r},t)>}{m} \quad (18)$$

is a locally averaged "potential energiality" of the ChWP;

$$\overline{<k(\vec{v},\vec{r},t)>} = \frac{\overline{<T(\vec{p},\vec{r},t)>}}{m} = \frac{1}{2m^2} \int_{-\infty}^{\infty} \rho(\vec{p},t)|\vec{p}|^2 d\vec{p} = \frac{1}{2} \int_{-\infty}^{\infty} \rho(\vec{v},t)|\vec{v}|^2 d\vec{v} \quad (19)$$

is the globally averaged "kinetic energiality" of the ChWP, where $\vec{v} := (v_x, v_y, v_z)$ is speed of the ChWP. We also introduce a notation: $|\vec{v}|^2 := v_x^2 + v_y^2 + v_z^2$,

$$\vec{v} := dv_x dv_y dv_z, \quad \int_{-\infty}^{\infty}\int_{-\infty}^{\infty}\int_{-\infty}^{\infty} dv_x dv_y dv_z := \int_{-\infty}^{\infty} d\vec{v}. \quad (20)$$

Substituting Exs. (16) – (19), taking into account the notation (20), into expression (15), we obtain a globally averaged "efficiency" of the ChWP, independent of a "mass" of the particle (21)

$$\overline{<s_r(t)>} = \int_{t_1}^{t_2} \left\{ \frac{1}{2} \int_{-\infty}^{\infty} \rho(\vec{v},t)|\vec{v}|^2 d\vec{v} + \int_{-\infty}^{\infty} \rho(\vec{r},t)[<u(\vec{r},t)> - <\varepsilon(\vec{r},t)>]d\vec{r} \right\} dt.$$

In one-dimensional consideration, Ex. (21) has the form (21a)

$$\overline{<s_x(t)>} = \int_{t_1}^{t_2} \left\{ \frac{1}{2} \int_{-\infty}^{\infty} \rho(v_x,t) v_x^2 dv_x + \int_{-\infty}^{\infty} \rho(x,t)[<u(x,t)> - <\varepsilon(x,t)>]dx \right\} dt,$$

we will call this expression «globally averaged *x*-efficiency of the ChWP».

## 2.3 Stationary state of the ChWP

Consider a stationary globally averaged state of the stochastic system under study, i.e., when the average ChWP behavior does not depend on time *t*.

In this case, a globally averaged efficiency of the ChWP (21) acquires a simplified form

$$\overline{<S_r>} = \int_{t_1}^{t_2} \left\{ \frac{1}{2} \int_{-\infty}^{\infty} \rho(\vec{v})|\vec{v}|^2 d\vec{v} + \int_{-\infty}^{\infty} \rho(\vec{r})[<u(\vec{r})> - <\varepsilon(\vec{r})>]d\vec{r} \right\} dt, \quad (22)$$

Let's represent expression (22) in a coordinate form. To do so, let's perform the following steps:

1]. We write PDF $\rho(\vec{r})$ as a product of two probability amplitude densities (PADs) $\psi(\vec{r})$:

$$\rho(\vec{r}) = \psi(\vec{r})\psi(\vec{r}) = \psi^2(\vec{r}). \quad (23)$$

2]. Let's use the coordinate representation of the average ChWP speed raised to the *n*-th power {see (A2.12) in Appendix 2}. In particular, for *n* = 2, we have

$$\overline{v_x^2} = \int_{-\infty}^{+\infty} \rho(v_x) v_x^2 dv_x = \int_{-\infty}^{+\infty} \psi(x)\left(\mp i\eta_x \frac{\partial}{\partial x}\right)^2 \psi(x) dx = -\eta_x^2 \int_{-\infty}^{\infty} \psi(x) \frac{\partial^2 \psi(x)}{\partial x^2} dx. \quad (24)$$

In the 3-dimensional case, expression (24) has the form (25)

$$\overline{|\vec{v}|^2} = \int_{-\infty}^{\infty} \rho(\vec{v})|\vec{v}|^2 d\vec{v} = \int_{-\infty}^{+\infty} \psi(\vec{r})(\mp i\eta_r \nabla)^2 \psi(\vec{r}) d\vec{r} = -\eta_r^2 \int_{-\infty}^{\infty} \psi(\vec{r}) \nabla^2 \psi(\vec{r}) dr,$$

where $\nabla^2 = \frac{\partial^2}{\partial x^2} + \frac{\partial^2}{\partial y^2} + \frac{\partial^2}{\partial z^2}$ is a Laplace operator;

$$\eta_r = \frac{2\sigma_r^2}{\tau_{rcor}} = const \quad (26)$$

is a scaling parameter, see (A2.12*a*) in Appendix 2;

$$\sigma_r = \frac{1}{\sqrt{3}}\sqrt{\sigma_x^2 + \sigma_y^2 + \sigma_z^2} \quad (27)$$

is a standard deviation of a random 3-dimensional trajectory of the ChWP from the conditional center of the considered stochastic system (see Figure 1);

$$\tau_{rcor} = \frac{1}{3}(\tau_{xcor} + \tau_{ycor} + \tau_{zcor}) \quad (28)$$

is an autocorrelation interval of a given 3-dimensional stationary random process. As could be seen from Figure 1 and Figure A1.1 in Appendix 1. The autocorrelation interval shows how smooth the random process under study is, i.e. how fast the ChWP moves and changes direction. The larger the autocorrelation interval, the lower the ChWP speed and the slower the change in the direction of its movement.

3]. Using expression (25), we represent the globally averaged kinetic energiality of the ChWP (19) as

$$\overline{< k(\vec{v},\vec{r}) >} = \frac{\overline{<T(\vec{p},\vec{r})>}}{m} = \frac{1}{2}\int_{-\infty}^{\infty}\rho(\vec{v})|\vec{v}|^2 d\vec{v} = -\frac{\eta_r^2}{2}\int_{-\infty}^{\infty}\psi(\vec{r})\nabla^2\psi(\vec{r})d\vec{r}. \quad (29)$$

4]. Substituting expressions (23) and (29) into the integral (22), we obtain the functional of the globally averaged efficiency of the ChWP in the coordinate representation                                                                                        (30)

$$\overline{< s_r >} = \int_{t_1}^{t_2}\int_{-\infty}^{\infty}\left(-\frac{\eta_r^2}{2}\psi(\vec{r})\nabla^2\psi(\vec{r}) + \psi^2(\vec{r})[< u(\vec{r}) > - < \varepsilon(\vec{r}) >]\right)d\vec{r}dt.$$

## 2.4 Derivation of the stationary stochastic Euler-Poisson equation

Let's find an equation for extremals $\psi(\vec{r})$ of the functional (30). Since there are no time-dependent functions in Ex. (30), we will look for a condition of extremality of the internal functional

$$w_r = \int_{-\infty}^{\infty}\left(-\frac{\eta_r^2}{2}\psi(\vec{r})\nabla^2\psi(\vec{r}) + \psi^2(\vec{r})[< u(\vec{r}) > - < \varepsilon(\vec{r}) >]\right)d\vec{r}. \quad (31)$$

We first consider an one-dimensional case, with functional (31) taking the form

$$w_x = \int_{-\infty}^{\infty}\left(-\frac{\eta_x^2}{2}\psi(x)\frac{\partial^2\psi(x)}{\partial x^2} + \psi^2(x)[< u(x) > - < \varepsilon(x) >]\right)dx. \quad (32)$$

where $$\eta_x = \frac{2\sigma_x^2}{\tau_{xcor}} = const. \quad (32a)$$

In the calculus of variations, it is shown [11,12] that the extremal $f(x)$ of the functional of the general form

$$I(f) = \int_{-\infty}^{\infty} L(x, f, f', f'', \ldots, f^{(k)})\,dx, \quad (33)$$

where $$f' := \frac{\partial f}{\partial x}, \quad f'' := \frac{\partial^2 f}{\partial x^2}, \quad \ldots, f^{(k)} := \frac{\partial^n f}{\partial x^n}$$

is determined by the generalized Euler-Lagrange equation [11,12]

$$\frac{\partial L}{\partial f} - \frac{\partial}{\partial x}\left(\frac{\partial L}{\partial f'}\right) + \frac{\partial^2}{\partial x^2}\left(\frac{\partial L}{\partial f''}\right) - \cdots + (-1)^k \frac{\partial^k}{\partial x^k}\left(\frac{\partial L}{\partial f^{(k)}}\right) = 0. \quad (34)$$

However, it should be noted that under other boundary conditions, the extremal of the functional (33) is determined by the Euler-Poisson-Elsgolts equation [11, P.P. 326 – 329, Elsgolts_LE]

$$\frac{\partial L}{\partial f} - \frac{\partial}{\partial x}\left\{\frac{\partial L}{\partial f'}\right\} + \frac{\partial^2}{\partial x^2}\left\{\frac{\partial L}{\partial f''}\right\} - \cdots + (-1)^k \frac{\partial^k}{\partial x^k}\left\{\frac{\partial L}{\partial f^{(k)}}\right\} = 0, \quad (34a)$$

where $\dfrac{\partial}{\partial x}\left\{\dfrac{\partial L}{\partial f'}\right\} = \dfrac{\partial}{\partial x}\left(\dfrac{\partial L}{\partial f'}\right) + \dfrac{\partial}{\partial f}\left(\dfrac{\partial L}{\partial f'}\right)\dfrac{\partial f}{\partial x} + \dfrac{\partial}{\partial f'}\left(\dfrac{\partial L}{\partial f'}\right)\dfrac{\partial f'}{\partial x} + \cdots + \dfrac{\partial}{\partial f^{(k)}}\left(\dfrac{\partial L}{\partial f'}\right)\dfrac{\partial f^{(k)}}{\partial x}$

is first total partial derivative with respect to $x$;

$\dfrac{\partial^2}{\partial x^2}\left\{\dfrac{\partial L}{\partial f''}\right\} = \dfrac{\partial^2}{\partial x^2}\left(\dfrac{\partial L}{\partial f''}\right) + \dfrac{\partial}{\partial f}\left(\dfrac{\partial L}{\partial f''}\right)\dfrac{\partial^2 f}{\partial x^2} + \dfrac{\partial}{\partial f'}\left(\dfrac{\partial L}{\partial f''}\right)\dfrac{\partial^2 f'}{\partial x^2} + \cdots + \dfrac{\partial}{\partial f^{(k)}}\left(\dfrac{\partial L}{\partial f''}\right)\dfrac{\partial^2 f^{(k)}}{\partial x_i^2}$

is the second total partial derivative with respect to $x$, and so on.

In the case of searching for an extremal $f = \psi(x)$ of Functional (32), we have the Lagrangian

$$L = -\frac{\eta_x^2}{2}\psi(x)\frac{\partial^2 \psi(x)}{\partial x^2} + \psi^2(x)[<u(x)> - <\varepsilon(x)>], \qquad (35)$$

depending only on $\psi$ and $\psi''$, while the generalized Euler-Lagrange equation (34) is simplified

$$\frac{\partial L}{\partial \psi} + \frac{\partial^2}{\partial x^2}\left(\frac{\partial L}{\partial \psi''}\right) = 0, \qquad (36)$$

because in this case, all other terms in equation (34) are equal to zero.

We write the terms from Eq. (36)

$$\frac{\partial L}{\partial \psi} = -\frac{\eta_x^2}{2}\frac{\partial^2 \psi(x)}{\partial x^2} + 2\psi(x)[<u(x)> - <\varepsilon(x)>], \qquad \frac{\partial L}{\partial \psi''} = -\frac{\eta_x^2}{2}\psi(x). \qquad (37)$$

Substituting Exs. (37) into Eq. (36), we obtain a one-dimensional stationary stochastic equation that does not depend on a particle mass

$$-\frac{\eta_x^2}{2}\frac{\partial^2 \psi(x)}{\partial x^2} + \psi(x)[<u(x)> - <\varepsilon(x)>] = 0. \qquad (38)$$

If instead of the Euler-Lagrange equation (34) we use the Euler-Poisson-Elsgolts equation (34$a$), then instead of Eq. (36) we get the equation

$$\frac{\partial L}{\partial \psi} + \frac{\partial^2}{\partial x^2}\left\{\frac{\partial L}{\partial \psi''}\right\} = 0, \qquad (38a)$$

where $\dfrac{\partial^2}{\partial x^2}\left\{\dfrac{\partial L}{\partial \psi''}\right\} = \dfrac{\partial^2}{\partial x^2}\left(\dfrac{\partial L}{\partial \psi''}\right) + \dfrac{\partial}{\partial \psi}\left(\dfrac{\partial L}{\partial \psi''}\right)\dfrac{\partial^2 \psi}{\partial x^2} + \dfrac{\partial}{\partial \partial \psi''}\left(\dfrac{\partial L}{\partial \psi''}\right)\dfrac{\partial^2 \partial \psi''}{\partial x^2}.$

Substituting Lagrangian (35) into Eq. (38$a$), in this case we obtain the following equation for extremals $\psi(x)$ of Functional (32)

$$-\frac{3\eta_x^2}{4}\frac{\partial^2 \psi(x)}{\partial x^2} + \psi(x)[<u(x)> - <\varepsilon(x)>] = 0. \qquad (38b)$$

Obviously, Eqs. (38) and (38b) differ from each other by a factor of 3/2 in front of the first term. Further research has shown that the Eq. (38b) is more preferable (see §2.9).

Consider now the 3-dimensional case. Let's represent Functional (31) in expanded form (39)

$$w = \int_{-\infty}^{\infty}\int_{-\infty}^{\infty}\int_{-\infty}^{\infty}\left(-\frac{\eta_r^2}{2}\psi(x,y,z)\nabla^2\psi(x,y,z) + \psi^2(x,y,z)[<u(x,y,z)> - <\varepsilon(x,y,z)>]\right)dxdydz.$$

For the 3-dimensional case, the general functional has the form

$$I[f] = \int_{-\infty}^{\infty}\int_{-\infty}^{\infty}\int_{-\infty}^{\infty} L(x_1, x_2, x_3, f, f_1, f_2, f_3, f_{11}, f_{22}, f_{33})dx_1 dx_2 dx_3, \quad (40)$$

where $\quad f_i := \frac{\partial f}{\partial x_i}, \quad f_{ii} := \frac{\partial^2 f}{\partial x_i \partial x_i} = \frac{\partial^2 f}{\partial x_i^2} \quad (i = 1,2,3).$ (41)

The extremal $f(x_1, x_2, x_3)$ of this functional is determined by the Euler-Poisson-Elsgolts equation [11, see P.P. 326 - 329, Elsgolts_LE] (42)

$$\frac{\partial L}{\partial f} - \frac{\partial}{\partial x_1}\left\{\frac{\partial L}{\partial f_1}\right\} - \frac{\partial}{\partial x_2}\left\{\frac{\partial L}{\partial f_2}\right\} - \frac{\partial}{\partial x_3}\left\{\frac{\partial L}{\partial f_3}\right\} + \frac{\partial^2}{\partial x_1^2}\left\{\frac{\partial L}{\partial f_{11}}\right\} + \frac{\partial^2}{\partial x_2^2}\left\{\frac{\partial L}{\partial f_{22}}\right\} + \frac{\partial^2}{\partial x_3^2}\left\{\frac{\partial L}{\partial f_{33}}\right\} = 0,$$

where (43)

$$\frac{\partial}{\partial x_i}\left\{\frac{\partial L}{\partial f_i}\right\} = \frac{\partial}{\partial x_i}\left(\frac{\partial L}{\partial f_i}\right) + \frac{\partial}{\partial f}\left(\frac{\partial L}{\partial f_i}\right)\frac{\partial f}{\partial x_i} + \frac{\partial}{\partial f_1}\left(\frac{\partial L}{\partial f_i}\right)\frac{\partial f_1}{\partial x_i} + \frac{\partial}{\partial f_2}\left(\frac{\partial L}{\partial f_i}\right)\frac{\partial f_2}{\partial x_i} + \frac{\partial}{\partial f_3}\left(\frac{\partial L}{\partial f_i}\right)\frac{\partial f_3}{\partial x_i} +$$

$$+ \frac{\partial}{\partial f_{11}}\left(\frac{\partial L}{\partial f_i}\right)\frac{\partial f_{11}}{\partial x_i} + \frac{\partial}{\partial f_{22}}\left(\frac{\partial L}{\partial f_i}\right)\frac{\partial f_{22}}{\partial x_1} + \frac{\partial}{\partial f_{33}}\left(\frac{\partial L}{\partial f_i}\right)\frac{\partial f_{33}}{\partial x_i}$$

is first total partial derivatives with the respect to $x_i$ ($i = 1,2,3$);

$$\frac{\partial^2}{\partial x_i^2}\left\{\frac{\partial L}{\partial f_{ii}}\right\} = \frac{\partial^2}{\partial x_i^2}\left(\frac{\partial L}{\partial f_{ii}}\right) + \frac{\partial}{\partial f}\left(\frac{\partial L}{\partial f_{ii}}\right)\frac{\partial^2 f}{\partial x_i^2} + \frac{\partial}{\partial f_1}\left(\frac{\partial L}{\partial f_{ii}}\right)\frac{\partial^2 f_1}{\partial x_i^2} + \frac{\partial}{\partial f_2}\left(\frac{\partial L}{\partial f_{ii}}\right)\frac{\partial^2 f_2}{\partial x_i^2} +$$

$$+ \frac{\partial}{\partial f_3}\left(\frac{\partial L}{\partial f_{ii}}\right)\frac{\partial^2 f_3}{\partial x_i^2} + \frac{\partial}{\partial f_{11}}\left(\frac{\partial L}{\partial f_{ii}}\right)\frac{\partial^2 f_{11}}{\partial x_i^2} + \frac{\partial}{\partial f_{22}}\left(\frac{\partial L}{\partial f_{ii}}\right)\frac{\partial^2 f_{22}}{\partial x_i^2} + \frac{\partial}{\partial f_{33}}\left(\frac{\partial L}{\partial f_{ii}}\right)\frac{\partial^2 f_{33}}{\partial x_i^2} \quad (44)$$

is second total partial derivatives with the respect to $x_i$ ($i = 1,2,3$).

In the case of the 3-dimensional functional (39), we have the Lagrangian:

$$L = -\frac{\eta_r^2}{2}\psi(x,y,z)\nabla^2\psi(x,y,z) + \psi^2(x,y,z)[<u(x,y,z)> - <\varepsilon(x,y,z)>], \quad (45)$$

wherein: $\quad x_1 = x, \quad x_2 = y, \quad x_3 = z; \quad f(x_1, x_2, x_3) = \psi(x, y, z),$

$\psi_x := \frac{\partial \psi}{\partial x}, \quad \psi_y := \frac{\partial \psi}{\partial y}, \quad \psi_z := \frac{\partial \psi}{\partial z}, \quad \psi_{xx} := \frac{\partial^2 \psi}{\partial x^2}, \quad \psi_{yy} := \frac{\partial^2 \psi}{\partial y^2}, \quad \psi_{zz} := \frac{\partial^2 \psi}{\partial z^2}.$

___

Since Lagrangian (45) depends only on $\psi$, $\psi_{xx}, \psi_{yy}$ and $\psi_{zz}$, the Euler-Poisson-Elsgolts equation (42) is simplified

$$\frac{\partial L}{\partial \psi} + \frac{\partial^2}{\partial x^2}\left\{\frac{\partial L}{\partial \psi_{xx}}\right\} + \frac{\partial^2}{\partial y^2}\left\{\frac{\partial L}{\partial \psi_{yy}}\right\} + \frac{\partial^2}{\partial z^2}\left\{\frac{\partial L}{\partial \psi_{zz}}\right\} = 0, \qquad (46)$$

since in this case all the other terms in Eq. (42) are equal to zero.

Taking into account Ex. (44), we write out the terms from Eq. (46)

$$\frac{\partial^2}{\partial x_i^2}\left\{\frac{\partial L}{\partial \psi_{ii}}\right\} = \frac{\partial^2}{\partial x_i^2}\left(\frac{\partial L}{\partial \psi_{ii}}\right) + \frac{\partial}{\partial \psi}\left(\frac{\partial L}{\partial \psi_{ii}}\right)\frac{\partial^2 \psi}{\partial x_i^2} + \frac{\partial}{\partial \psi_{xx}}\left(\frac{\partial L}{\partial \psi_{ii}}\right)\frac{\partial^2 \psi_{xx}}{\partial x_i^2} + \qquad (46a)$$

$$+ \frac{\partial}{\partial \psi_{yy}}\left(\frac{\partial L}{\partial \psi_{ii}}\right)\frac{\partial^2 \psi_{yy}}{\partial x_i^2} + \frac{\partial}{\partial \psi_{zz}}\left(\frac{\partial L}{\partial \psi_{ii}}\right)\frac{\partial^2 \psi_{zz}}{\partial x_i^2}, \quad (\text{where } i = 1,2,3),$$

which are simplified to the expression

$$\frac{\partial^2}{\partial x_i^2}\left\{\frac{\partial L}{\partial \psi_{ii}}\right\} = \frac{\partial^2}{\partial x_i^2}\left(\frac{\partial L}{\partial \psi_{ii}}\right) + \frac{\partial}{\partial \psi}\left(\frac{\partial L}{\partial \psi_{ii}}\right)\frac{\partial^2 \psi}{\partial x_i^2} \qquad (46b)$$

because the remaining terms in (46a) are equal to zero.

We substitute the Lagrangian (45) into Eq. (46), and preliminarily write out the terms from this equation, taking into an account Ex. (46b)

$$\frac{\partial L}{\partial \psi} = -\frac{\eta_r^2}{2}\left(\frac{\partial^2 \psi(x,y,z)}{\partial x^2} + \frac{\partial^2 \psi(x,y,z)}{\partial y^2} + \frac{\partial^2 \psi(x,y,z)}{\partial z^2}\right) + 2\psi(x,y,z) \times$$
$$\times [< u(x,y,z) > - < \varepsilon(x,y,z) >], \qquad (47)$$

$$\frac{\partial^2}{\partial x^2}\left\{\frac{\partial L}{\partial \psi_{xx}}\right\} = -\frac{\eta_r^2}{2}\frac{\partial^2 \psi(x,y,z)}{\partial x^2} + \left(-\frac{\eta_r^2}{2}\frac{\partial^2 \psi(x,y,z)}{\partial x^2}\right) = -2\frac{\eta_r^2}{2}\frac{\partial^2 \psi(x,y,z)}{\partial x^2},$$

$$\frac{\partial^2}{\partial y^2}\left\{\frac{\partial L}{\partial \psi_{yy}}\right\} = -\frac{\eta_r^2}{2}\frac{\partial^2 \psi(x,y,z)}{\partial y^2} + \left(-\frac{\eta_r^2}{2}\frac{\partial^2 \psi(x,y,z)}{\partial x^2}\right) = -2\frac{\eta_r^2}{2}\frac{\partial^2 \psi(x,y,z)}{\partial y^2},$$

$$\frac{\partial^2}{\partial y^2}\left\{\frac{\partial L}{\partial \psi_{zz}}\right\} = -\frac{\eta_r^2}{2}\frac{\partial^2 \psi(x,y,z)}{\partial z^2} + \left(-\frac{\eta_r^2}{2}\frac{\partial^2 \psi(x,y,z)}{\partial z^2}\right) = -2\frac{\eta_r^2}{2}\frac{\partial^2 \psi(x,y,z)}{\partial z^2}.$$

As a result of substituting Exs. (47) into the Euler-Poisson-Elsgolts equation (46), we obtain the following three-dimensional stochastic equation for finding extremals $\psi(x,y,z) = \psi(\vec{r})$ of the functional of a globally averaged efficiency of the ChWP (39) $\qquad\qquad (48)$

$$-\frac{3\eta_r^2}{2}\left\{\frac{\partial^2 \psi(x,y,z)}{\partial x^2} + \frac{\partial^2 \psi(x,y,z)}{\partial y^2} + \frac{\partial^2 \psi(x,y,z)}{\partial z^2}\right\} + 2[< u(x,y,z) > - < \varepsilon(x,y,z) >]\psi(x,y,z) = 0.$$

Multiply both sides of this equation by –1, and write it in a compact form

$$\frac{3\eta_{\vec{r}}^2}{4}\nabla^2\psi(\vec{r}) + [<\varepsilon(\vec{r})> - <u(\vec{r})>]\,\psi(\vec{r}) = 0, \tag{49}$$

Eq. (49) will be referred to as "*the mass-independent stationary (i.e., time - independent) three-dimensional stochastic Euler-Poisson equation*" (or abbreviated as the "*stationary stochastic Euler–Poisson equation*).

## 2.5 Stationary stochastic Schrödinger-Euler-Poisson equation

Let's consider the case when the locally averaged mechanical energiality of the ChWP $<\varepsilon(\vec{r})>$ is constant at all points of the stochastic system under study

$$<\varepsilon(\vec{r})> = \varepsilon = const. \tag{50}$$

In such a stationary stochastic system, a randomly changing kinetic energiality of the ChWP $k(x,y,z,t)$ and its potential energiality $u(x,y,z,t)$ transform into each other in such a way that their averaged sum at each point $\vec{r} := (x, y, z)$ is constant

$$<k(\vec{r})> + <u(\vec{r})> = \varepsilon = const. \tag{51}$$

In this case, stochastic equation (49) takes the form

$$-\frac{3\eta_{\vec{r}}^2}{4}\nabla^2\psi(\vec{r}) + <u(\vec{r})>\psi(\vec{r}) = \varepsilon\,\psi(\vec{r}). \tag{52}$$

We compare this equation with the stationary Schrödinger equation

$$-\frac{\hbar^2}{2m}\nabla^2\psi(\vec{r}) + U(\vec{r})\psi(\vec{r}) = E\psi(\vec{r}), \tag{53}$$

where $\hbar$ is the reduced Planck constant ($\hbar = 1.055 \times 10^{-34}$ J/Hz).

First, we divide both parts of the Schrödinger equation (53) by the particle mass $m$

$$-\frac{\hbar^2}{2m^2}\nabla^2\psi(\vec{r}) + \frac{U(\vec{r})}{m}\psi(\vec{r}) = \frac{E}{m}\psi(\vec{r}). \tag{54}$$

We write Eq. (54) in the following form

$$-\frac{1}{2}\left(\frac{\hbar}{m}\right)^2\nabla^2\psi(\vec{r}) + u(\vec{r})\psi(\vec{r}) = \varepsilon\psi(\vec{r}), \tag{55}$$

where, according to the terminology introduced above (16) – (19):

$\varepsilon = E/m$ is the "mechanical energiality" of the particle; (56)

$u(\vec{r}) = U(\vec{r})/m$ is the "potential energiality" of the particle. (57)

Comparing Eqs. (52) and (55), we find that for

$$\frac{\hbar}{m} = \sqrt{\frac{3}{2}}\eta_r = \sqrt{\frac{3}{2}}\frac{2\sigma_r^2}{\tau_{rcor}}\left(\frac{m^2}{s}\right) \quad \text{and} \quad u(\vec{r}) = <u(\vec{r})> \quad (58)$$

these equations completely coincide.

*The ratio of the volume of a cylinder to the volume of a sphere inscribed in it is **3/2**. Archimedes was so shocked by this discovery that he requested his kinsmen to engrave a sphere inscribed in a cylinder on his tombstone. It is believed that later Cicero found the grave of Archimedes thanks to this symbol (note by S. Petukhov).* 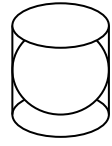

Thus, it turned out that the time-independent Schrödinger equation (53) is a special case of the stochastic equation (52).

Therefore, Eq. (52) will be referred to as "*the mass-independent stationary three-dimensional stochastic Schrödinger-Euler-Poisson equation*" (or abbreviated, "*stationary stochastic Schrödinger-Euler-Poisson equation*").

Due to the smallness of the reduced Planck constant $\hbar$, the requirement $\sqrt{3/2}\,\eta_r = \hbar/m$ applies to a microscopic stochastic systems, or to an extremely precise macrolevel systems, such as the LIGO project.

Meanwhile, in all previous considerations, no restrictions were imposed on the scale parameter $\eta_r = 2\sigma_r^2/\tau_{rcor}$ (see Appendix 1 and Appendix 2), that is, the scale parameter $\eta_r$ can depend on a standard deviation $\sigma_r$ and an autocorrelation interval $\tau_{rcor}$ of a random process that involved ChWP of any scale and any quality. In other words, in contrast to the Schrödinger equation (53), stochastic equations (49) and (52) are suitable for describing stationary states of any compact objects or subjects, randomly wandering in 3-dimensional space (real or phase) in physics, biology, sociology, psychology and economics, etc.

In contrast to Eq. (53), in fact, which was hypothesized by Erwin Schrödinger in 1926, stochastic Eq.s (52) and (49) are obtained on the basis of the variational principle. That is, stochastic Eq.s (52) and (49) are the conditions for searching for extremals of the globally averaged "efficiency" functional of ChWP (39), which is symmetric with respect to the simultaneous tendency of any stochastic system to "order" and "chaos". This is the additional significance of the result obtained.

## 2.6 Time-dependent stochastic Euler-Poisson equation

Let the averaged characteristics of the random trajectory of the ChWP movement change with time, but so slowly that in each small time interval $\Delta t$, all these characteristics can be considered unchanged. Such an unstable behavior of a particle is considered in Appendix 1 and is called a pseudo-stationary random process (PSRP).

Let's also assume that for such a pseudo-stationary stochastic system (i.e., the nonequilibrium state of the ChWP), its globally averaged mechanical energiality changes insignificantly over time

$$\overline{<\varepsilon(x,y,z,t)>} = \overline{<k(x,y,z,t)>} + \overline{<u(x,y,z,t)>}.$$

Since the change in this globally averaged mechanical energiality is slow, we can write

$$\overline{<\varepsilon(x,y,z,t)>} = \overline{<\varepsilon(x,y,z,t_0)>} \pm \overline{<\varepsilon_k(x,y,z,t)>}, \quad (59)$$

where $\overline{<\varepsilon(x,y,z,t_0)>}$ is the initial value of the globally averaged mechanical energiality of the ChWP at the time $t_0$;

$\pm \overline{<\varepsilon_k(x,y,z,t)>}$ is a small change {increase (+) or decrease (–)} of the globally averaged mechanical energiality of the ChWP for a time interval $t$, associated with a change in its kinetic energiality $\Delta k$.

To reduce the calculations, consider the one-dimensional case, without loss of generality of conclusions for the case of 3-dimensions, and represent Ex. (59) in an abbreviated form

$$\overline{<\varepsilon(x,t)>} = \overline{<\varepsilon(x,t_0)>} \pm \overline{<\varepsilon_k(x,t)>}. \quad (60)$$

Let's write Ex. (60) as follows

$$\overline{<\varepsilon(x,t)>} = \int_{-\infty}^{\infty} \rho(x,t)<\varepsilon(x,t_0)>dx \pm \int_{-\infty}^{\infty} \rho(\varepsilon_{kx},t)\varepsilon_{kx}d\varepsilon_{kx}, \quad (61)$$

where $\varepsilon_{kx}$ is a local change in the mechanical energiality of the ChWP due to a
    slight change in its kinetic energiality in the direction of the *X*-axis (see
    § A2.2 in Appendix 2);

$\rho(\varepsilon_{kx}, t)$ is the probability distribution function (PDF) of changes of the mechanical energiality of the ChWP $\varepsilon_{kx}$.

Let's substitute Ex. (61) into the globally averaged *x*-efficiency of the ChWP (21*a*)

$$\overline{<s_x(t)>} = \int_{t_1}^{t_2} \left\{ \frac{1}{2} \int_{-\infty}^{\infty} \rho(v_x,t) v_x^2 dv_x + \int_{-\infty}^{\infty} \rho(x,t)[<u(x,t)> - <\varepsilon(x,t_0)>]dx \pm \right.$$
$$\left. \pm \int_{-\infty}^{\infty} \rho(\varepsilon_{kx},t)\varepsilon_{kx} d\varepsilon_{kx} \right\} dt. \qquad (62)$$

Let's write Ex. (62) in the coordinate representation. For this, we express the PDFs $\rho(x,t)$, $\rho(v_x,t)$ and $\rho(\varepsilon_x,t)$ in terms of the probability amplitude $\psi(x,t)$.

According to Ex. (A1.47) in Appendix 1

$$\rho(x,t) = \psi(x,t)\psi(x,t) = \psi^2(x,t), \qquad (63)$$

According to Ex. (A2.13) in Appendix 2

$$\frac{1}{2}\overline{v_x^2(t)} = \frac{1}{2}\int_{-\infty}^{\infty} \rho(v_x,t) v_x^2 dv_x = -\frac{\eta_x^2}{2}\int_{-\infty}^{\infty} \psi(x,t) \frac{\partial^2 \psi(x,t)}{\partial x^2} dx, \qquad (64)$$

where {see Ex. (A2.13*a*) in Appendix 2}

$$\eta_x = \frac{2\sigma_x^2(t-t_0)}{\tau_{xкор}(t-t_0)} \approx \frac{2\sigma_x^2}{\tau_{xкор}} = const$$

is a constant scale parameter*, with dimensions (m²/s).

*The invariability of the scale parameter $\eta_x = \frac{2\sigma_x^2(t)}{\tau_{xcor}(t)} \approx$ constant with a change in the wave function $\psi(x,t)$ in this case is due to the fact that the variance $\sigma_x^2(t)$ and the autocorrelation interval $\tau_{xcor}(t)$ of the stochastic system under study change (i.e., increase or decrease) practically simultaneously and in the same way.*

*Keeping unchanged the ratio of the main averaged characteristics of the investigated random process $\sigma_x^2(t)/\tau_{xcor}(t)$ in Appendix 2 is called: "The law of proportional constancy of the scale parameter $\eta_x$ of the stochastic system." This effect is similar to the phenomenon described in the molecular kinetic theory, when the expansion of a gas is invariably accompanied by a slowdown in the speed of motion of its atoms and / or molecules and a decrease in the intensity of their collisions (i.e., a change in the direction of their motion); and, conversely, gas compression is accompanied by an increase in the speed and intensity of collisions of its atoms and / or molecules.*

According to Ex. (A2.31) in Appendix 2

$$\overline{<\varepsilon_k(x,t)>} = \int_{-\infty}^{\infty} \rho(\varepsilon_{kx},t)\varepsilon_{kx} d\varepsilon_{kx} = \pm i\frac{\eta_x^2}{D} \int_{-\infty}^{+\infty} \psi(x,t) \frac{\partial \psi(x,t)}{\partial t} dx, \qquad (65)$$

where *D* is the imaginary part of the complex self-diffusion coefficient $B = iD$ of a chaotically wandering particle (ChWP) with dimension m$^2$/s, {see Ex. (A2.19) in Appendix 2}. It is assumed that $D$ = const.

As well as

$$\int_{-\infty}^{\infty} \rho(x,t) <u(x,t)> dx = \int_{-\infty}^{\infty} \psi^2(x,t) <u(x,t)> dx, \qquad (66)$$

$$\int_{-\infty}^{\infty} \rho(x,t)\{<\varepsilon(x,t_0)> dx = \int_{-\infty}^{\infty} \psi^2(x,t) <\varepsilon(x,t_0)> dx. \qquad (67)$$

Substituting Ex.s (63) – (67) into integral (62), we obtain the coordinate representation of the pseudo-stationary globally averaged *x*-efficiency of the ChWP

$$\overline{<s_x(t)>} = \int_{t_1}^{t_2} \int_{-\infty}^{\infty} \left( -\frac{\eta_x^2}{2} \psi(x,t) \frac{\partial^2 \psi(x,t)}{\partial x^2} + \psi^2(x,t)[<u(x,t)> - <\varepsilon(x,t_0)>] \pm \right.$$

$$\left. \pm i \frac{\eta_x^2}{D} \psi(x,t) \frac{\partial \psi(x,t)}{\partial t} \right) dx dt. \qquad (68)$$

In the case of 3-dimensional consideration, this functional has the form

$$\overline{<s_r(t)>} = \int_{t_1}^{t_2} \int_{-\infty}^{\infty} \int_{-\infty}^{\infty} \int_{-\infty}^{\infty} \begin{pmatrix} -\frac{\eta_r^2}{2} \psi(\vec{r},t) \nabla^2 \psi(\vec{r}) + [<u(\vec{r},t)> - \\ -<\varepsilon(\vec{r},t_0)>]\psi^2(\vec{r},t) \pm i\frac{\eta_r^2}{D}\psi(\vec{r},t)\frac{\partial \psi(\vec{r},t)}{\partial t} \end{pmatrix} dx dy dz dt,$$

where $\vec{r} := (x, y, z)$. \hfill (69)

Let's find the equation for the extremals $\psi(x,y,z,t)$ of Functional (69).

First, recall that the extremality condition for a functional of the form

$$I = \int_{x_{01}}^{x_{02}} \int_{-\infty}^{\infty} \int_{-\infty}^{\infty} \int_{-\infty}^{\infty} L(x_0, x_1, x_2, x_3, f, f_0, f_1, f_2, f_3, f_{00}, f_{11}, f_{22}, f_{33}) dx_0 dx_1 dx_2 dx_3$$

$$\text{where} \quad f_i := \frac{\partial f}{\partial x_i}, \quad f_{ii} := \frac{\partial^2 f}{\partial x_i \partial x_i} = \frac{\partial^2 f}{\partial x_i^2} \quad (i = 0,1,2,3) \qquad (70)$$

is determined by the Euler-Poisson-Elsgolts equation [11, see P.P. 326, The Calculus Of Variations: L. Elsgolts]

$$\frac{\partial L}{\partial f} - \frac{\partial}{\partial x_0}\left\{\frac{\partial L}{\partial f_0}\right\} - \frac{\partial}{\partial x_1}\left\{\frac{\partial L}{\partial f_1}\right\} - \frac{\partial}{\partial x_2}\left\{\frac{\partial L}{\partial f_2}\right\} - \frac{\partial}{\partial x_3}\left\{\frac{\partial L}{\partial f_3}\right\} + \frac{\partial^2}{\partial x_0^2}\left\{\frac{\partial L}{\partial f_{00}}\right\} + \frac{\partial^2}{\partial x_1^2}\left\{\frac{\partial L}{\partial f_{11}}\right\} +$$

$$+ \frac{\partial^2}{\partial x_2^2}\left\{\frac{\partial L}{\partial f_{22}}\right\} + \frac{\partial^2}{\partial x_3^2}\left\{\frac{\partial L}{\partial f_{33}}\right\} = 0, \qquad (71)$$

where \hfill (72)

$$\frac{\partial}{\partial x_i}\left\{\frac{\partial L}{\partial f_i}\right\} = \frac{\partial}{\partial x_i}\left(\frac{\partial L}{\partial f_i}\right) + \frac{\partial}{\partial f}\left(\frac{\partial L}{\partial f_i}\right)\frac{\partial f}{\partial x_i} + \frac{\partial}{\partial f_0}\left(\frac{\partial L}{\partial f_i}\right)\frac{\partial f_0}{\partial x_i} + \frac{\partial}{\partial f_1}\left(\frac{\partial L}{\partial f_i}\right)\frac{\partial f_1}{\partial x_i} + \frac{\partial}{\partial f_2}\left(\frac{\partial L}{\partial f_i}\right)\frac{\partial f_2}{\partial x_i} +$$

$$+ \frac{\partial}{\partial f_3}\left(\frac{\partial L}{\partial f_i}\right)\frac{\partial f_3}{\partial x_i} + \frac{\partial}{\partial f_{00}}\left(\frac{\partial L}{\partial f_i}\right)\frac{\partial f_{00}}{\partial x_i} + \frac{\partial}{\partial f_{11}}\left(\frac{\partial L}{\partial f_i}\right)\frac{\partial f_{11}}{\partial x_i} + \frac{\partial}{\partial f_{22}}\left(\frac{\partial L}{\partial f_i}\right)\frac{\partial f_{22}}{\partial x_1} + \frac{\partial}{\partial f_{33}}\left(\frac{\partial L}{\partial f_i}\right)\frac{\partial f_{33}}{\partial x_i}$$

is the first complete partial derivatives with respect to $x_i$ ($i = 0,1,2,3$);

$$\frac{\partial^2}{\partial x_i^2}\left\{\frac{\partial L}{\partial f_{ii}}\right\} = \frac{\partial^2}{\partial x_i^2}\left(\frac{\partial L}{\partial f_{ii}}\right) + \frac{\partial}{\partial f}\left(\frac{\partial L}{\partial f_{ii}}\right)\frac{\partial^2 f}{\partial x_i^2} + \frac{\partial}{\partial f_0}\left(\frac{\partial L}{\partial f_{ii}}\right)\frac{\partial^2 f_0}{\partial x_i^2} + \frac{\partial}{\partial f_1}\left(\frac{\partial L}{\partial f_{ii}}\right)\frac{\partial^2 f_1}{\partial x_i^2} +$$

$$+ \frac{\partial}{\partial f_2}\left(\frac{\partial L}{\partial f_{ii}}\right)\frac{\partial^2 f_2}{\partial x_i^2} + \frac{\partial}{\partial f_3}\left(\frac{\partial L}{\partial f_{ii}}\right)\frac{\partial^2 f_3}{\partial x_i^2} + \frac{\partial}{\partial f_{00}}\left(\frac{\partial L}{\partial f_{ii}}\right)\frac{\partial^2 f_{00}}{\partial x_i^2} + \frac{\partial}{\partial f_{11}}\left(\frac{\partial L}{\partial f_{ii}}\right)\frac{\partial^2 f_{11}}{\partial x_i^2} +$$

$$+ \frac{\partial}{\partial f_{22}}\left(\frac{\partial L}{\partial f_{ii}}\right)\frac{\partial^2 f_{22}}{\partial x_i^2} + \frac{\partial}{\partial f_{33}}\left(\frac{\partial L}{\partial f_{ii}}\right)\frac{\partial^2 f_{33}}{\partial x_i^2} \qquad (73)$$

is the second complete partial derivatives with respect to $x_i$ ($i = 0,1,2,3$);

In the case of functional (69), we have Lanrangian:

$$L = -\frac{\eta_r^2}{2}\psi(x,y,z,t)\left(\frac{\partial^2\psi(x,y,z,t)}{\partial x^2} + \frac{\partial^2\psi(x,y,z,t)}{\partial y^2} + \frac{\partial^2\psi(x,y,z,t)}{\partial z^2}\right) +$$
$$+[<u(x,y,z,t)> - <\varepsilon(x,y,z,t)>]\psi^2(x,y,z,t) \pm i\frac{\eta_r^2}{D}\psi(x,y,z,t)\frac{\partial\psi(x,y,z,t)}{\partial t}, \quad (74)$$

wherein

$$x_1 = t, \quad x_1 = x, \quad x_2 = y, \quad x_3 = z; \quad f(x_0, x_1, x_2, x_3) = \psi(x,y,z,t), \quad (75)$$

$$\psi_t := \frac{\partial\psi}{\partial t}, \quad \psi_x := \frac{\partial\psi}{\partial x}, \quad \psi_y := \frac{\partial\psi}{\partial y}, \quad \psi_z := \frac{\partial\psi}{\partial z},$$

$$\psi_{tt} := \frac{\partial^2\psi}{\partial t^2}, \quad \psi_{xx} := \frac{\partial^2\psi}{\partial x^2}, \quad \psi_{yy} := \frac{\partial^2\psi}{\partial y^2}, \quad \psi_{zz} := \frac{\partial^2\psi}{\partial z^2}.$$

Since Lagrangian (74) depends only on $\psi$, $\psi_t, \psi_{xx}, \psi_{yy}$ and $\psi_{zz}$, the Euler-Poisson-Elsgolts equation (71) is simplified

$$\frac{\partial L}{\partial \psi} - \frac{\partial}{\partial t}\left\{\frac{\partial L}{\partial \psi_t}\right\} + \frac{\partial^2}{\partial x^2}\left\{\frac{\partial L}{\partial \psi_{xx}}\right\} + \frac{\partial^2}{\partial y^2}\left\{\frac{\partial L}{\partial \psi_{yy}}\right\} + \frac{\partial^2}{\partial z^2}\left\{\frac{\partial L}{\partial \psi_{zz}}\right\} = 0. \qquad (76)$$

because all other terms in (71) are equal to zero.

As a result of substituting the Lagrangian (74) into expressions (72) and (73), taking into account the notation (75), we obtain the following terms in equation (76)

$$\frac{\partial L}{\partial \psi} = -\frac{\eta_r^2}{2}\frac{\partial^2\psi(\vec{r},t)}{\partial x^2} + 2\psi(\vec{r},t)[<u(\vec{r},t)> - <\varepsilon(\vec{r},t_0)>] \pm i\frac{\eta_r^2}{D}\frac{\partial\psi(\vec{r},t)}{\partial t},$$

$$\frac{\partial}{\partial t}\left\{\frac{\partial L}{\partial \psi_t}\right\} = \pm i2\frac{\eta_r^2}{D}\frac{\partial\psi(\vec{r},t)}{\partial t}, \qquad \frac{\partial^2}{\partial x^2}\left\{\frac{\partial L}{\partial \psi_{xx}}\right\} = -2\frac{\eta_r^2}{2}\frac{\partial^2\psi(\vec{r},t)}{\partial x^2}, \qquad (76a)$$

$$\frac{\partial^2}{\partial y^2}\left\{\frac{\partial L}{\partial \psi_{yy}}\right\} = -2\frac{\eta_r^2}{2}\frac{\partial^2\psi(\vec{r},t)}{\partial x^2}, \qquad \frac{\partial^2}{\partial z^2}\left\{\frac{\partial L}{\partial \psi_{zz}}\right\} = -2\frac{\eta_r^2}{2}\frac{\partial^2\psi(\vec{r},t)}{\partial x^2}.$$

Substituting these expressions into the Euler-Poisson-Elsgolts (76), we obtain the desired mass-independent three-dimensional stochastic equation for determining the extremals $\psi(\vec{r},t) = \psi(x,y,z,t)$ of the functional of the globally averaged ChWP efficiency (69)

$$\pm i \frac{\eta_r^2}{D} \frac{\partial \psi(\vec{r},t)}{\partial t} = -\frac{3\eta_r^2}{2} \nabla^2 \psi(\vec{r},t) + 2[<u(\vec{r},t)> - <\varepsilon(\vec{r},t_0)>] \psi(\vec{r},t). \quad (77)$$

Eq. (67) is a system of two equations

$$\begin{cases} -i \frac{\eta_r^2}{2D} \frac{\partial \psi(\vec{r},t)}{\partial t} = -\frac{3\eta_r^2}{4} \nabla^2 \psi(\vec{r},t) + [<u(\vec{r},t)> - <\varepsilon(\vec{r},t_0)>]\psi(\vec{r},t), & (77a) \\ i \frac{\eta_r^2}{2D} \frac{\partial \psi(\vec{r},t)}{\partial t} = -\frac{3\eta_r^2}{4} \nabla^2 \psi(\vec{r},t) + [<u(\vec{r},t)> - <\varepsilon(\vec{r},t_0)>]\psi(\vec{r},t). & (77b) \end{cases}$$

This suggests that in a slowly varying (i.e., pseudo-stationary) stochastic system, two processes should occur simultaneously, described by the two Eq.s (77a) and (77b).

Eq. (67) will be called "*the mass-independent pseudo-stationary (i.e., time-dependent) three-dimensional stochastic Euler-Poisson equation*" (or, in short, "*the time-dependent stochastic Euler-Poisson equation*").

## 2.7 Time-dependent stochastic Schrödinger-Euler-Poisson equation

Let's suppose that at the initial moment of time $t_0$ the globally averaged mechanical energiality of the ChWP is zero (i.e., $<\varepsilon(\vec{r},t_0)> = 0$), then Eq. (77) takes the form

$$\pm i \frac{\eta_r^2}{2D} \frac{\partial \psi(\vec{r},t)}{\partial t} = -\frac{3}{4} \eta_r^2 \nabla^2 \psi(\vec{r},t) + <u(\vec{r},t)> \psi(\vec{r},t). \quad (78)$$

In the case when $D = \eta_r$ we obtain the self-consistent stochastic Euler-Poisson equation

$$\pm i \frac{\eta_r}{2} \frac{\partial \psi(\vec{r},t)}{\partial t} = -\frac{3}{4} \eta_r^2 \nabla^2 \psi(\vec{r},t) + <u(\vec{r},t)> \psi(\vec{r},t). \quad (78a)$$

Let's compare Eq. (78a) with the time-dependent Schrödinger equation

$$i\hbar \frac{\partial \psi(\vec{r},t)}{\partial t} = -\frac{\hbar^2}{2m} \nabla^2 \psi(\vec{r},t) + U(\vec{r},t) \psi(\vec{r},t). \quad (79)$$

First, we divide both sides of the Eq. (79) by the particle mass $m$, as a result, taking into account definition (56) and (57), we obtain

$$i\frac{\hbar}{m}\frac{\partial \psi(\vec{r},t)}{\partial t} = -\frac{1}{2}\frac{\hbar^2}{m^2}\nabla^2\psi(\vec{r},t) + u(\vec{r},t)\,\psi(\vec{r},t), \qquad (80)$$

where $u(\vec{r},t) = U(\vec{r},t)/m$ is the potential energiality of the ChWP.

Obviously, for $<u(\vec{r})> = u(\vec{r})$ and $\frac{\hbar}{m} = \eta_r = \frac{2\sigma_r^2}{\tau_{rcor}}$ self-consistent stochastic Eq. (78a) and Schrödinger equation (80) coincide up to constant coefficients and the ± sign.

*It is interesting to note that Erwin Schrödinger wrote Eq. (4'') in "Quantisierung als Eigenwertproblem, Vierte Mitteilung", Annalen der Physik (1926) [1] in the following form*

$$\Delta\psi - \frac{8\pi^2}{h^2}V\psi \pm \frac{4\pi i}{h}\frac{\partial\psi}{\partial t} = 0. \qquad (81)$$

*Let's rearrange the terms in this expression and take into account that $\hbar = h/2\pi$,*

$$\pm i\hbar\frac{\partial\psi}{\partial t} = -\frac{1}{2}\hbar^2\nabla^2\psi + V\psi. \qquad (82)$$

*In this case, there is a complete analogy between the original Schrödinger equation (82) and the stochastic Eq. (78a).*

Therefore, equation (78a) will be called: "the mass-independent non-stationary three-dimensional stochastic Schrödinger-Euler-Poisson equation" (or, in short, the "time-dependent stochastic Schrödinger-Euler-Poisson equation").

In contrast to the time-dependent Schrödinger equation (79), the non-stationary stochastic Schrödinger-Euler-Poisson equation (78a) is applicable to the study of time-varying averaged states of ChWP of any scale.

## 2.8 The stochastic equation of imaginary self-diffusion

If in the stochastic Eq. (78a) we equate to zero the locally averaged potential energiality (i.e., $<u(\vec{r},t)> = 0$), then it takes the form of the stochastic equation of imaginary self-diffusion (which is a special case of the Fokker-Planck-Kolmogorov equation)

$$\frac{\partial\psi(\vec{r},t)}{\partial t} = \mp i\frac{3}{2}D\nabla^2\psi(\vec{r},t) \qquad (83)$$

with an complex self-diffusion coefficient $B = i\frac{3}{2}D$.

Thus, the proposed model of a stochastic system presupposes the possibility of studying various variants of the ChWP behavior, depending on its averaged characteristics.

## 2.9 The problem of factor 3

In the absence of the dependence of the wave function on time, i.e. for $\partial\psi(\vec{r},t)/\partial t = 0$ and $<\varepsilon(\vec{r},t_0)>=\varepsilon$, Eq. (77) becomes the stationary stochastic Schrödinger-Poisson-Euler equation (41)

$$-\frac{3\eta_r^2}{4}\nabla^2\psi(\vec{r})+<u(\vec{r})>\psi(\vec{r})=\varepsilon\psi(\vec{r}), \qquad (84)$$

which is to be expected.

However, on the other hand, one would expect that the equation of imaginary self-diffusion (83) should have the form

$$\frac{\partial\psi(\vec{r},t)}{\partial t} = \mp iD\nabla^2\psi(\vec{r},t), \qquad (85)$$

since the complex self-diffusion coefficient was determined as $B = iD$ {see Ex. (A2.19) in Appendix 2}.

That is, if instead of factor 3 in the numerator of the first term in equation (84) there were factor 2, then Eq. (52) {or (84)} would have the more preferable form

$$-\frac{\eta_r^2}{2}\nabla^2\psi(\vec{r})+<u(\vec{r})>\psi(\vec{r})=\varepsilon\psi(\vec{r}) . \qquad (86)$$

In this case, Eq. (78) would also be more concise

$$\pm i\frac{\eta_r^2}{2D}\frac{\partial\psi(\vec{r},t)}{\partial t} = -\frac{\eta_r^2}{2}\nabla^2\psi(\vec{r},t)+<u(\vec{r},t)>\psi(\vec{r},t), \qquad (87)$$

and for $\partial\psi(\vec{r},t)/\partial t = 0$ and $<u(\vec{r})> = 0$, which becomes Eq. (85).

The ambiguity of the situation lies in the fact that if the Euler-Lagrange equation

$$\frac{\partial L}{\partial\psi} + \frac{\partial^2}{\partial x^2}\left(\frac{\partial L}{\partial\psi_{xx}}\right) + \frac{\partial^2}{\partial y^2}\left(\frac{\partial L}{\partial\psi_{yy}}\right) + \frac{\partial^2}{\partial z^2}\left(\frac{\partial L}{\partial\psi_{zz}}\right) = 0, \qquad (88)$$

___

were used instead of the Euler-Poisson equation (46), then it is easy to make sure that the substitution into this equation of the Lagrangian (45)

$$L = -\frac{\eta_r^2}{2}\psi(x,y,z)\nabla^2\psi(x,y,z) + \psi^2(x,y,z)[<u(x,y,z)> - <\varepsilon(x,y,z)>]$$

leads to the expected result, i.e. to obtain equation

$$-\frac{\eta_r^2}{2}\nabla^2\psi(\vec{r}) + <u(\vec{r})>\psi(\vec{r}) = <\varepsilon(\vec{r})>\psi(\vec{r}). \qquad (89)$$

It would seem that using the Euler-Lagrange equation (87) is a more correct way, because in this equation the factor 3 disappears.

But if instead of the Euler-Poisson-Elsgolts equation (76), we use the Euler-Lagrange equation

$$\frac{\partial L}{\partial \psi} - \frac{\partial}{\partial t}\left(\frac{\partial L}{\partial \psi_t}\right) + \frac{\partial^2}{\partial x^2}\left(\frac{\partial L}{\partial \psi_{xx}}\right) + \frac{\partial^2}{\partial y^2}\left(\frac{\partial L}{\partial \psi_{yy}}\right) + \frac{\partial^2}{\partial z^2}\left(\frac{\partial L}{\partial \psi_{zz}}\right) = 0, \qquad (90)$$

then after substituting into it the Lagrangian (74)

$$L = -\frac{\eta_r^2}{2}\psi(\vec{r},t)\nabla^2\psi(\vec{r},t) + [<u(\vec{r},t)> - <\varepsilon(\vec{r},t)>]\psi^2(\vec{r},t) \pm i\frac{\eta_r^2}{D}\psi(\vec{r},t)\frac{\partial\psi(\vec{r},t)}{\partial t}$$

the following equation is obtained

$$-\frac{\eta_x^2}{2}\frac{\partial^2\psi(\vec{r},t)}{\partial x^2} + \psi(\vec{r},t)[<u(\vec{r},t)> - <\varepsilon(\vec{r},t_0)>] = 0, \qquad (91)$$

in which the first time derivative of the wave function ($\frac{\partial\psi(\vec{r},t)}{\partial t}$) is absent, and this equation does not correspond to the time-dependent Schrödinger equation (79).

We do not know which formalism is more correct: of Euler-Lagrange or Euler-Poisson. We can only state the facts:

- using the Euler-Lagrange formalism, the derivation of the time-independent Schrödinger equation is obtained, in which there is no multiplier (factor) 3. But this formalism does not allow the derivation of the time-dependent Schrödinger equation;

- when using the Euler-Poisson formalism, the derivation of the time-independent Schrödinger equation and the time-dependent Schrödinger equation is obtained up to coefficients. But in this case, there is a factor 3, not the expected factor of 2.

Let's formulate the essence of the problem using the example of a 2-dimensional functional of general form

$$I[f] = \int_{-\infty}^{\infty}\int_{-\infty}^{\infty} L(x_0, x_1, f, f_0, f_1, f_{00}, f_{01}, f_{11}, \ldots, f_{11\ldots1}) dx_1 dx_2, \quad (92)$$

where $\quad f_i = \frac{\partial f}{\partial x_i}, \quad f_{ij} = \frac{\partial^2 f}{\partial x_i \partial x_j}, \quad f_{iij} = \frac{\partial^3 f}{\partial x_i \partial x_i \partial x_j}, \quad \ldots \quad (i = 0,1; \; j = 0,1)$.

The extremals of this functional can satisfy the Euler-Lagrange equation [32, 33]

$$\frac{\partial L}{\partial f} - \frac{\partial}{\partial x_0}\left(\frac{\partial L}{\partial f_0}\right) - \frac{\partial}{\partial x_1}\left(\frac{\partial L}{\partial f_1}\right) + \frac{\partial^2}{\partial x_0^2}\left(\frac{\partial L}{\partial f_{00}}\right) + \frac{\partial^2}{\partial x_0 \partial x_1}\left(\frac{\partial L}{\partial f_{01}}\right) + \frac{\partial^2}{\partial x_1^2}\left(\frac{\partial L}{\partial f_{11}}\right) - \cdots +$$

$$+ (-1)^n \frac{\partial^n}{\partial x_1^n}\left(\frac{\partial L}{\partial f_{11\ldots1}}\right) = 0, \quad (93)$$

or the Euler-Poisson-Elsgolts equation [11]

$$\frac{\partial L}{\partial f} - \frac{\partial}{\partial x_0}\left\{\frac{\partial L}{\partial f_0}\right\} - \frac{\partial}{\partial x_1}\left\{\frac{\partial L}{\partial f_1}\right\} + \frac{\partial^2}{\partial x_0^2}\left\{\frac{\partial L}{\partial f_{00}}\right\} + \frac{\partial^2}{\partial x_0 \partial x_1}\left\{\frac{\partial L}{\partial f_{01}}\right\} + \frac{\partial^2}{\partial x_1^2}\left\{\frac{\partial L}{\partial f_{11}}\right\} - \cdots +$$

$$+ (-1)^n \frac{\partial^n}{\partial x_1^n}\left\{\frac{\partial L}{\partial f_{11\ldots1}}\right\} = 0, \quad (94)$$

where

$$\frac{\partial}{\partial x_i}\left\{\frac{\partial L}{\partial f_i}\right\} = \frac{\partial}{\partial x_i}\left(\frac{\partial L}{\partial f_i}\right) + \frac{\partial}{\partial f}\left(\frac{\partial L}{\partial f_i}\right)\frac{\partial f}{\partial x_i} + \frac{\partial}{\partial f_0}\left(\frac{\partial L}{\partial f_i}\right)\frac{\partial f_0}{\partial x_i} + \frac{\partial}{\partial f_1}\left(\frac{\partial L}{\partial f_i}\right)\frac{\partial f_1}{\partial x_i} +$$

$$+ \frac{\partial}{\partial f_{00}}\left(\frac{\partial L}{\partial f_i}\right)\frac{\partial f_{00}}{\partial x_i} + \frac{\partial}{\partial f_{11}}\left(\frac{\partial L}{\partial f_i}\right)\frac{\partial f_{11}}{\partial x_i} + \frac{\partial}{\partial f_{10}}\left(\frac{\partial L}{\partial f_i}\right)\frac{\partial f_{10}}{\partial x_i} + \cdots \quad (95)$$

is the first full partial derivatives with respect to $x_i$ ($i = 0,1$).

$$\frac{\partial^2}{\partial x_i^2}\left\{\frac{\partial L}{\partial f_{ii}}\right\} = \frac{\partial^2}{\partial x_i^2}\left(\frac{\partial L}{\partial f_{ii}}\right) + \frac{\partial}{\partial f}\left(\frac{\partial L}{\partial f_{ii}}\right)\frac{\partial^2 f}{\partial x_i^2} + \frac{\partial}{\partial f_0}\left(\frac{\partial L}{\partial f_{ii}}\right)\frac{\partial^2 f_0}{\partial x_i^2} + \frac{\partial}{\partial f_1}\left(\frac{\partial L}{\partial f_{ii}}\right)\frac{\partial^2 f_1}{\partial x_i^2} +$$

$$+ \frac{\partial}{\partial f_{00}}\left(\frac{\partial L}{\partial f_{ii}}\right)\frac{\partial^2 f_{00}}{\partial x_i^2} + \frac{\partial}{\partial f_{11}}\left(\frac{\partial L}{\partial f_{ii}}\right)\frac{\partial^2 f_{11}}{\partial x_i^2} + \frac{\partial}{\partial f_{01}}\left(\frac{\partial L}{\partial f_{ii}}\right)\frac{\partial^2 f_{01}}{\partial x_i^2} + \cdots \quad (96)$$

is the second full partial derivatives with respect to $x_i$ ($i = 0,1$) $\quad (97)$

$$\frac{\partial}{\partial x_i \partial x_j}\left\{\frac{\partial L}{\partial f_{ij}}\right\} = \frac{\partial^2}{\partial x_i \partial x_j}\left(\frac{\partial L}{\partial f_{ij}}\right) + \frac{\partial}{\partial f}\left(\frac{\partial L}{\partial f_{ij}}\right)\frac{\partial^2 f}{\partial x_i \partial x_j} + \frac{\partial}{\partial f_0}\left(\frac{\partial L}{\partial f_{ij}}\right)\frac{\partial^2 f_0}{\partial x_i \partial x_j} +$$

$$+ \frac{\partial}{\partial f_1}\left(\frac{\partial L}{\partial f_{ij}}\right)\frac{\partial^2 f_1}{\partial x_i \partial x_j} + \frac{\partial}{\partial f_{00}}\left(\frac{\partial L}{\partial f_{ij}}\right)\frac{\partial^2 f_{00}}{\partial x_i \partial x_j} + \frac{\partial}{\partial f_{11}}\left(\frac{\partial L}{\partial f_{ij}}\right)\frac{\partial^2 f_{11}}{\partial x_i \partial x_j} + \frac{\partial}{\partial f_{10}}\left(\frac{\partial L}{\partial f_{ij}}\right)\frac{\partial^2 f_{10}}{\partial x_i \partial x_j} + \cdots$$

is second mixed complete partial derivatives with respect to $x_i$ & $x_j$ ($i = 0,1; \; j = 0,1$).

For the author, the Euler-Poisson-Elsgolts formalism looks more preferable, since leads to the derivation of both Schrödinger equations up to constant coefficients. However, the question (more precisely "the problem of factor 3") remains open.

## 2.10 The stochastic quantum operators

Let's show how operators are obtained in mass-independent stochastic quantum mechanics (MSQM). To do this, let us return to considering the model of a chaotically wandering particle (ChWP) shown in Figure 1.

During the chaotic movement of a particle in the vicinity of the conditional center, it constantly changes the direction of its movement. Therefore, a particle at each moment of time has an angular momentum

$$\vec{L} = \vec{r} \times \vec{p}, \qquad (98)$$

where $\vec{r}$ is the radius vector from the conditional center to the particle (Figure 1); $\vec{p} = m\vec{v}$ is the instantaneous value and direction of the particle momentum vector.

Let's divide both sides of the vector Ex. (98) by the value $m$, as a result we obtain the angular velocity vector

$$\vec{\omega} = \frac{\vec{L}}{m|\vec{r}|^2} = \frac{\vec{r} \times \vec{v}}{|\vec{r}|^2}. \qquad (99)$$

We represent the vector equation (99) in the component form

$$\omega_x = \frac{1}{|\vec{r}|^2}(yv_z - zv_y), \quad \omega_y = \frac{1}{|\vec{r}|^2}(zv_x - xv_z), \quad \omega_z = \frac{1}{|\vec{r}|^2}(xv_y - yv_x). \qquad (100)$$

Let's average these components

$$\overline{\omega_x} = \frac{1}{|\vec{r}|^2}(y\overline{v_z} - z\overline{v_y}), \quad \overline{\omega_y} = \frac{1}{|\vec{r}|^2}(z\overline{v_x} - x\overline{v_z}), \quad \overline{\omega_z} = \frac{1}{|\vec{r}|^2}(x\overline{v_y} - y\overline{v_x}). \qquad (101)$$

We use the coordinate representation of the averaged components of the velocity vector (A2.2) in Appendix 2 for $n = 1$

$$\overline{v_x} = \int_{-\infty}^{+\infty} \psi(x)\left(\pm i\eta_x \frac{\partial}{\partial x}\right)\psi(x)dx = \left(\pm \frac{i\eta_x}{2}\frac{\partial}{\partial x}\right)\int_{-\infty}^{+\infty} \psi(x)\psi(x)dx, \qquad (102)$$

$$\overline{v_y} = \int_{-\infty}^{+\infty} \psi(y)\left(\pm i\eta_y \frac{\partial}{\partial y}\right)\psi(y)dy = \left(\pm \frac{i\eta_x}{2}\frac{\partial}{\partial y}\right)\int_{-\infty}^{+\infty} \psi(y)\psi(y)dy, \qquad (103)$$

$$\overline{v_z} = \int_{-\infty}^{+\infty} \psi(z)\left(\pm i\eta_z \frac{\partial}{\partial z}\right)\psi(z)dz = \left(\pm \frac{i\eta_x}{2}\frac{\partial}{\partial z}\right)\int_{-\infty}^{+\infty} \psi(z)\psi(z)dz. \qquad (104)$$

Let's prove that these expressions are true using the example of Ex. (102).

First, let's show that the expression

$$\overline{v_x} = \int_{-\infty}^{+\infty} \psi(x)\left(\pm i\eta_x \frac{\partial}{\partial x}\right)\psi(x)dx = \int_{-\infty}^{+\infty} \pm \frac{i\eta_x}{2}\frac{\partial}{\partial x}[\psi(x)\psi(x]\,dx, \qquad (105)$$

is true, since the equality holds

$$-\frac{i\eta_x}{2}\frac{\partial}{\partial x}[\psi(x)\psi(x] = -\frac{i\eta_x}{2}\left[\frac{\partial \psi(x)}{\partial x}\psi(x) + \psi(x)\frac{\partial \psi(x)}{\partial x}\right] = -\frac{i\eta_x}{2}\left[2\psi(x)\frac{\partial \psi(x)}{\partial x}\right] =$$
$$= -i\eta_x\left[\psi(x)\frac{\partial \psi(x)}{\partial x}\right] = \psi(x)\left(-i\eta_x\frac{\partial}{\partial x}\right)\psi(x).$$

Due to the fact that the operations of integration and differentiation are commutative, we finally write

$$\overline{v_x} = \int_{-\infty}^{+\infty}\psi(x)\left(-i\eta_x\frac{\partial}{\partial x}\right)\psi(x)dx = \int_{-\infty}^{+\infty} -\frac{i\eta_x}{2}\frac{\partial}{\partial x}[\psi(x)\psi(x]dx = \left(-\frac{i\eta_x}{2}\frac{\partial}{\partial x}\right)\int_{-\infty}^{+\infty}\psi(x)\psi(x)dx,$$

this is what it was required to prove (Q.E.D).

Take into account that, for example, in (101)

$$\int_{-\infty}^{\infty} \psi(x)\psi(x)dx = \int_{-\infty}^{\infty} \rho(x)dx = 1 \qquad (106)$$

this, in fact, means that a wandering particle randomly moves in one direction or the other, so that the average values of the components of its velocity (102) – (104) are equal to zero (i.e., $\overline{v_x} = 0$, $\overline{v_y} = 0$ and $\overline{v_z} = 0$).

Therefore, identities (102) – (104) are equivalent to mass-independent stochastic operators of the components of the velocity vector

$$\hat{v}_x = \mp \frac{\eta_r}{i}\frac{\partial}{\partial x}, \qquad \hat{v}_y = \mp \frac{\eta_r}{i}\frac{\partial}{\partial y}, \qquad \hat{v}_z = \mp \frac{\eta_r}{i}\frac{\partial}{\partial z}, \qquad (107)$$

here it is taken into account that for the isotropic case $\eta_x = \eta_y = \eta_z = \eta_r$.

The mass-independent stochastic operators (107), for $\frac{\hbar}{m} = \eta_r$, correspond to the operators of the components of the QM momentum vector [22]

$$\hat{p}_x = \frac{\hbar}{i}\frac{\partial}{\partial x}, \qquad \hat{p}_y = \frac{\hbar}{i}\frac{\partial}{\partial y}, \qquad \hat{p}_z = \frac{\hbar}{i}\frac{\partial}{\partial z}.$$

Substituting expressions (102) – (104) into expressions (101), taking into account (107), we obtain mass-independent stochastic operators of the components of the ChWP angular velocity vector

$$\hat{\omega}_x = \mp \frac{\eta_r}{i|\vec{r}|^2}\left(y\frac{\partial}{\partial z} - z\frac{\partial}{\partial y}\right), \quad (108)$$

$$\hat{\omega}_y = \mp \frac{\eta_r}{i|\vec{r}|^2}\left(z\frac{\partial}{\partial x} - x\frac{\partial}{\partial z}\right),$$

$$\hat{\omega}_z = \mp \frac{\eta_r}{i|\vec{r}|^2}\left(x\frac{\partial}{\partial y} - y\frac{\partial}{\partial x}\right),$$

which correspond to the quantum mechanical operators of the components of the angular momentum vector [22]

$$\hat{L}_x = \frac{\hbar}{i}\left(y\frac{\partial}{\partial z} - z\frac{\partial}{\partial y}\right), \quad \hat{L}_y = \frac{\hbar}{i}\left(z\frac{\partial}{\partial x} - x\frac{\partial}{\partial z}\right), \quad \hat{L}_z = \frac{\hbar}{i}\left(x\frac{\partial}{\partial y} - y\frac{\partial}{\partial x}\right).$$

In a spherical coordinate system, stochastic operators (108) have the form

$$\hat{\omega}_x = \mp \frac{\eta_r}{i|\vec{r}|^2}\left(sin\,\phi\,\frac{\partial}{\partial \theta} - ctg\theta\,cos\,\phi\,\frac{\partial}{\partial \phi}\right), \quad (109)$$

$$\hat{\omega}_y = \mp \frac{\eta_r}{i|\vec{r}|^2}\left(cos\,\phi\,\frac{\partial}{\partial \theta} - ctg\theta\,sin\,\phi\,\frac{\partial}{\partial \phi}\right),$$

$$\hat{\omega}_z = \mp \frac{\eta_r}{i|\vec{r}|^2}\frac{\partial}{\partial \phi}.$$

The stochastic mass-independent operator of the square of the modulus of the angular velocity of the ChWP is

$$\hat{\omega}^2 = \hat{\omega}_x^2 + \hat{\omega}_y^2 + \hat{\omega}_z^2 = -\frac{\eta_r^2}{|\vec{r}|^4}\nabla^2_{\theta,\phi},$$

where

$$\nabla^2_{\theta,\phi} = \frac{1}{sin\,\theta}\frac{\partial}{\partial \theta}\left(sin\,\theta\,\frac{\partial}{\partial \theta}\right) + \frac{1}{sin^2\,\theta}\frac{\partial^2}{\partial \phi^2}. \quad (110)$$

All mass-independent stochastic quantum operators, analogous to the QM operators, can be obtained in a similar way. Only in the MSQM, instead of the ratio $\hbar/m$, there is a scale parameter $\eta_r$ (26) {or (A2.12a) in Appendix 1}, therefore, the BSCM is suitable for describing stochastic processes of any scale.

In a similar way, the mathematical apparatus of the entire mass-independent stochastic quantum mechanics (MSQM) can be built, which almost completely coincides with the mathematical apparatus of the QM. But MSQM is based on the principles of "ordinary" (classical) logic, and is suitable for describing quantum systems and effects of any scale.

## 2.10 The uncertainty principle in MSQM

The uncertainty in the velocity of a chaotically wandering particle (ChWP) is determined by the standard deviation

$$\sqrt{\overline{v_x^2}} = \sqrt{\int_{-\infty}^{+\infty} \psi(x)\left(\pm i\eta_x \frac{\partial}{\partial x}\right)^2 \psi(x)dx} = i\eta_x \sqrt{\int_{-\infty}^{+\infty} \psi(x)\frac{\partial^2 \psi(x)}{\partial x^2}dx}, \quad (112)$$

and the uncertainty in the particle coordinate is determined by the volatility

$$\sqrt{\overline{x^2}} = \sqrt{\int_{-\infty}^{+\infty} \psi(x)x^2\psi(x)dx}. \quad (113)$$

The joint uncertainty in coordinate and momentum can be represented as

$$\sqrt{\overline{v_x^2 x^2}} = \sqrt{\int_{-\infty}^{+\infty} \psi(x)\left(\pm i\eta_x \frac{\partial}{\partial x}\right)^2 x^2\psi(x)dx} = \sqrt{-\eta_x^2 \int_{-\infty}^{+\infty} \psi(x)\frac{\partial^2 x^2}{\partial x^2}\psi(x)dx} = \sqrt{2}i\eta_x \quad (114)$$

This uncertainty principle of the MSQM is equivalent to the Heisenberg's uncertainty principle $\Delta x \Delta p_x \geq 2\pi\hbar$.

## 3 CONCLUSIONS

The article proposes to bring attention to the fact that all the objects and subjects around us simultaneously strive for two goals equal in importance, but mutually opposite (i. e., antisymmetric):

- to the maximum possible order, which is expressed in the "principle of least action", and

- to the maximum possible chaos, which is expressed in the "principle of maximum entropy".

In other words, in the reality around us there is not a single completely deterministic or completely random entity. Everything is subject to the simultaneous striving for change and orderliness.

In order to simultaneously take into account both of these tendencies, this work proposes a unified "principle of the average efficiency extremum", in which both competing concepts, "order" and "chaos", coexist.

The article considers the averaged states of a particle (i.e., a compact solid body) of any size, which, under the influence of fluctuations in the environment and/or various long-range forces, continuously wanders (oscillates, displaces) in 3-dimensional space like a Brownian particle.

A constantly trembling (shifting, oscillating) body is represented as a chaotically wandering particle (ChWP) with a continuous trajectory of motion and volume. At the same time, the internal structure of the ChWP is not considered, as well as rotation and deformations of its shape are not taken into account.

These ChWP include the **centers of**: a valence electron in a hydrogen-like atom, a vibrating atom in the crystal lattice, the trembling yolk in a chicken egg, a floating moth in the vicinity of a burning lamp etc. (see the beginning of § 2.1).

All these stochastic systems are similar to each other and obey the same laws, taking into account different types of friction coefficient and viscosity of the medium surrounding the ChWP, as well as the duration of the average period of its behavior. For example, in order to average the chaotic flights of a bird in a cage, a week of continuous observation is required; while averaging the chaotic displacements of the galactic nucleus relative to the main line of its motion in outer space will require millions of years of research. But the results of such observations may turn out to be similar, despite the large difference in the scale of these events.

For example, in in Appendix 3 (or in §3.6 [27, arXiv:1702.01880]), it is predicted theoretically that the possible averaged states of the vibrating nucleus of a biological cell are similar to discrete states of a 3-dimensional quantum mechanical oscillator (i.e., an elementary particle under similar conditions). If these microscopic quantum effects are confirmed experimentally, then we will be able to outline ways to solve the measurement problem in stochastic quantum mechanics.

Within the framework of the mass-independent stochastic quantum mechanics (MSQM), the problem of studying "pure" states of pico-particles is proposed to be solved as follows. It is necessary to find (or simulate) a stochastic macroscopic system, similar to the investigated pico-scopic system (i.e., a chaotically wandering pico-particle), and carry out experiments with the macroscopic system without exerting a tangible effect on it. Then, the results of measurements at the macro level are projected onto possible similar manifestations of the picoscopic system.

Within the MSQM, such an approach to the study of "pure" states of a picoscopic and megascopic systems is possible, since the philosophical foundations of this stochastic mechanics are rooted in antiquity and are based on the belief that all levels of the Universe are similar to each other. In this sense, MSQM is a universal theory for all levels of organization of chaotically oscillating (shifting, trembling, wandering, moving) matter.

As applied to pico-particles (i.e., particles of atomic and subatomic scale), the MSQM corresponds to the stochastic quantum mechanics (SQM) of Edward Nelson [2]. In this case, the MSQM Eq.s (52) and (78$a$), derived in this article on the basis of the "principle of averaged efficiency extremum" of the ChWP, coincided with the corresponding Schrödinger Eq.s (53) and (79) up to coefficients.

In other words, in the mass-independent stochastic quantum mechanics (MSQM), the "pure" wave function $\psi(x,t)$ is the extremal of the functional of the averaged "efficiency" of the ChWP, written in the coordinate representation.

Thus, the stochastic equations (49), (52), (77), (78$a$), and (83) obtained in the article are the conditions for finding the extremals $\psi(x,t)$ of the functional of the globally averaged "efficiency" of a chaotically wandering particle (ChWP).

It is important to note that this stochastic functional is balanced in relation to the simultaneous striving of any stochastic system immediately to two mutually opposite (antisymmetric) goals:

- to "order" (that is, to determinism with the least energy losses), and
- to "chaos" (i.e., to the maximum entropy, or to the greatest uncertainty).

Stochastic Eq.s (52) and (78*a*) have a number of the following advantages over the corresponding Schrödinger equations (53) and (79):

1]. In the reasoning given to derive stochastic Eq.s (52) and (78*a*), no restrictions were imposed on a chaotically wandering particle (ChWP), except for the total energiality balances (51) and (59). That is, ChWP is an ordinary particle that has: volume, trajectory of movement, location and momentum at every moment of time. In other words, the derivation of the stochastic Schrödinger-Euler-Poisson Eq.s (52) and (78*a*), was obtained on the basis of "ordinary" (classical) logic using the theory of probability, the theory of generalized functions, and the calculus of variations.

Whereas in 95 years, since the appearance of Schrödinger's equations in 1926, many researchers have proposed various methods of deriving them, relying on the axioms of many different interpretations of quantum mechanics, but no universally recognized result has been obtained.

The scientific community has not succeeded in developing logically consistent justifications for the QM axioms. One of the reasons for the general dissatisfaction was the lack of a "beautiful" derivation of the Schrödinger equations.

2]. The reduced Planck's constant ($\hbar = 1.055 \times 10^{-34}$ J/Hz) limits the scope of the Schrödinger equations (53) and (79), and the entire QM as a whole, to the description of atomic and subatomic scale phenomena.

The fact is that the ratio $\hbar/m$, which is explicitly or latently present in the Schrödinger equations, only then turns out to be physically significant when the particle mass *m* is very small (for example, it is believed that the electron rest mass $m_e = 9.109 \times 10^{-31}$ kg).

Whereas the field of application of the stochastic Schrödinger-Euler-Poisson equations (52) and (78*a*) is not limited by anything.

To use Eq.s (52) and (78*a*) to describe the averaged states of any of the above stochastic systems, it is necessary to estimate their scale parameter $\eta_r$ (26). For this, it is necessary to determine the standard deviation $\sigma_r$ and the autocorrelation interval $\tau_{r\,cor}$ of a three-dimensional random process, in which the corresponding particle participates, on the basis of sufficiently long observations of the geometric center of the ChWP.

As an example, in in Appendix 3 shows the possibility of using the mass-independent stationary stochastic Schrödinger-Euler-Poisson equation (52) to obtain quantum numbers characterizing the possible averaged states of a chaotically oscillating nucleus of a biological cell during the interphase period.

3]. The stochastic equation 52) is also applicable to describe the averaged states of a chaotically moving geometric center of an electron in the vicinity of the nucleus of a hydrogen-like atom. If, as a result of statistical processing of indirect observations of the chaotic behavior of a valence electron in such an atom, it turns out that its scale parameter is

$$\eta_{er} = \frac{2\sigma_{er}^2}{\tau_{ercor}} \approx \frac{\hbar}{m_e} = \frac{1{,}055 \times 10^{-34}}{9{,}1 \times 10^{-31}} \approx 0{,}116 \times 10^{-3}\,\frac{m^2}{s},$$

then Eq.s (52) and (53) for this case will turn out to be almost completely equivalent. In this sense, the time-independent Schrödinger equation (53) can be regarded as a particular case of the stationary stochastic Schrödinger-Euler-Poisson equation (52).

4]. In Schrödinger's equations (53) and (79), the mass of an elementary particle is present. But this mass cannot be directly measured by macroscopic measuring instruments.

On the other hand, in the stochastic Schrödinger-Euler-Poisson equations (52) and (78$a$) there is no particle mass. In this case, the standard deviation $\sigma_r$ and the autocorrelation interval $\tau_{r\,cor}$ of a three-dimensional random process, in which the ChWP is involved, can always be estimated based on the statistical processing of the results of sufficiently long observations of practically any stochastic system. Therefore, the stochastic Eq.s (49), (52), (77), (78$a$), and (83) obtained in this article are of a universal nature.

5]. It has been introduced into the minds of several generations of physicists that an electron is a point-particle (that is, an entity that has no volume). This happened mainly because a point charge, according to Coulomb's law, must explode due to the infinite repulsive force of different parts of a very small charged particle.

There were also other arguments, for example, L.D. Landau believed that if the electron were not point-particle, then in strong collisions with other particles such stresses would arise in it that it would inevitably disintegrate into parts, but in practice this does not happen.

At the same time, the idea that elementary particles are point-particles leads to many problems (in particular, to ultraviolet divergences). In addition, the ordinary mind will never agree with the lack of volume in a real particle.

Therefore, attempts to get away from the point like nature of elementary particles were always. For example, renormalization and regulation (in particular, vacuum polarization around a point charge and charge diffusion due to fluctuations) form the illusion of particle volume, and many string theorists have moved from point to linear objects in a multidimensional space with 6 compactly folded dimensions.

Another thing is that quantum mechanics (QM) was originally formulated in such a way that it operates with wave functions, and not with particles. Therefore, for the QM methodology it does not matter whether an elementary particle is a point object or not. However, in a number of excrements at the time of registration, elementary particles manifest themselves as clearly local formations, rather than wave functions, diffuse throughout the Universe. The followers of Niels Bohr bypassed this problem with the help of the so-called instant "state reduction". Many experts do not like the fact that this process must proceed at an infinite speed, so the adherents of the many-worlds interpretation of Hugh Everett found it more acceptable to assume that it is not the wave function of a particle dispersed throughout the world that instantly collapses into a point, but the whole world instantly goes into one of the many possible states corresponding to the result of the experiment.

At the same time, according to a number of scientists (also modern), the concept of puffiness of elementary particles should be replaced by the concept of their "geometric center". Although the "geometric center" of any object is a mental construction, it does not contradict common sense, and the study of the dynamics of the "geometric center" of a complex object in some cases simplifies poorly formalized tasks.

In the proposed article, it is the geometric center of the ChWP that moves along a chaotic trajectory, despite the fact that the shape of the ChWP can change, oscillate and rotate.

6]. If we assume that the velocity and coordinate of a chaotically wandering particle (ChWP) are completely independent (i.e., uncorrelated) quantities, then a Brownian (and not quantum) particle is obtained, described by the diffusion equation (a particular case of the Fokker-Planck-Kolmagorov equation). For such a particle, the dispersion of its location increases with time. This is a model of the Markov process.

However, if a colloidal particle turns out to be in a small closed space, or in the field of action of a potential force, then the Markov character of the process is violated and in a number of cases the averaged behavior of the ChWP becomes quantum (i.e., its averaged states are quantized).

The theory of dynamic chaos deals with such processes, where, on the one hand, completely deterministic systems of differential equations with certain external and internal parameters can lead to pseudo-chaotic trajectories diverging in the Lyapunov fashion. On the other hand, in dynamic chaos, order periodically manifests itself, i.e. discrete ordered and/or periodic configurations (patterns).

"True" quantum indeterminism, according to the author, is due to the fact that the average state of any dynamical system is determined by the dominance of two fundamental factors: the "principle of least action" and "the principle of maximum entropy", i.e. the parity between the «order» and «chaos».

Despite the fact that the entropy of a distributed system is not mentioned in this article, nevertheless, it is latently present. Schrödinger-Euler-Poisson equations obtained in this article on the basis of the principle of ""principle of averaged efficiency extremum", lead to solutions (i.e., to the squares of the modulus of wave functions), which are simultaneously extremals of both the functional of the globally averaged efficiency and the Shannon entropy functional.

At the beginning of the method for studying a stochastic distributed system proposed in this article, we abstracted from chaos by means of averaging. However, the averaged order turned out to correspond to the maximum of entropy (uncertainty), as a guarantee of stable equilibrium of the considered distributed system.

7]. MSQM predicts that many stationary random processes (in which ChWP are involved) have the possibility of transition from one stationary state to another with the absorption or release of a certain part of the total mechanical energiality.

This is easy to check, for example, in the case of a moth constantly chaotically flying around a luminous lamp. With a video camera, you can record his chaotic movements for a long time. If you then scroll through the video recording at high speed, then the moth will not be visible on the screen, but there will be a stable blurry dark spot, which reflects the PDF of the location of its geometric center. It should be expected that if the moth is not disturbed by anything, then the blurred spot will resemble a Gaussian PDF with the greatest darkening in the area of the center of the light bulb. However, if the moth is somehow energetically influenced, for example, by heat or ultrasound with a certain frequency, then its average behavior can abruptly change. In this case, the blurred spot can change the configuration to the average shape of a ring or figure-eight, etc.

Also, the geometric center of a flower, depending on the intensity of gusts of wind, can, on average, describe a straight segment, a circle, an ellipse, a figure-eight, or another Lissajous figure.

Similar 2-D and 3-D quantum effects appear in all ChWP of any scale. This contains the main idea of mass-independent stochastic quantum mechanics (MSQM): "Studying stochastic objects of the macrocosm using conventional (benchtop) methods, we simultaneously obtain information about all similar objects of the microcosm and objects of cosmic scale.

The approach proposed in this paper makes it possible to derive the equations of nonrelativistic mass-independent stochastic quantum mechanics (MSQM) (37), (41), (67), (68) and (73) based on principles fundamentally different from the ideological foundations of modern QM interpretations: Copenhagen, Many-worlds, Consistent histories, Decoherence, etc., but the mathematical apparatus of the MSQM turns out to be completely analogous to the mathematical apparatus of the QM.

Apparently, many other equations of quantum field theory can be obtained in a similar way, for example: the Klein-Fock-Gordon equation, the Dirac equation, the Maxwell equations, etc. It is possible that the algorithm for deriving them is similar to the approach given in this work:

1) the average energy balance condition of the stochastic system is recorded;

2) mass is extracted from the average energy balance condition and the "efficiency" of this system is obtained;

3) to execute globally average the "efficiency" of the stochastic system under consideration;

4) all the averaged terms in the integrand of the globally averaged "efficiency" to represented through the PDF $\rho(x)$ and PDF $\rho(p_x)$;

5) all the terms of the Lagrangian of the globally averaged "efficiency" of the system are converted into a coordinate representation;

6) the equation for the extremals of the resulting functional is determined by means of the calculus of variations (i.e., using the Euler - Poisson equation).

It is possible that further research will confirm the validity of this approach to the derivation of the field theory equations.

We hope that this work will assist in the discovery and study of quantum phenomena not only of the microcosm, but also of the macro- and mega Worlds.

## 4 ACKNOWLEDGEMENTS

It is impossible to overestimate the help of the American mathematician David Reid (USA, Israel). The solution of a number of mathematical problems and the development of terminology took place in close cooperation with D. Reid.

I am grateful to David Reid (USA, Israel), Ph.D. V. A. Lukyanov, Ph.D. G.K. Tolokonnikov and M. Habbash for valuable remarks and comments made during the preparation of this article. I also express gratitude to my mentors, Dr. A.A. Kuznetsov and Dr. A.I. Kozlov. A great help in the discussion of this article was provided by Dr. A.A. Rukhadze, Dr. A.M. Ignatov and Dr. G.I. Shipov. Thanks for support to S.V. Musanov, mathematician V.P. Khramikhin, Dr. E.P. Mouser, Ph.D. T.S. Levy, A.N. Maslov, A.Yu. Bolotov L.A, Ph.D. S.V. Mizin (P.N. Lebedev Physical Institute of the Russian Academy of Sciences). Batanova and Carlos J. Rojas (Editor of the journal Avances en Ciencias e Ingeniería, Chile). The comments of Dr. L. S. F. Olavo (University of Brazil) and Dr. D. Edwards (Georgia State University, USA) were helpful to the author. The criticism, comments and suggestions of two reviewers of the journals "Entropy" and "Symmetry" (Basel, Switzerland), who, unfortunately, wished to remain anonymous, were important. The last revised version of the article was written on the recommendation of Prof. Fabio Rinaldi from Guglielmo Marconi University (Rome, Italy).

Appendix 1

## A1 Determination of the PDF of the derivative of a stationary and pseudo-stationary differentiated random process

Consider several realizations of the random process $\xi(t)$ (see Figure A1.1).

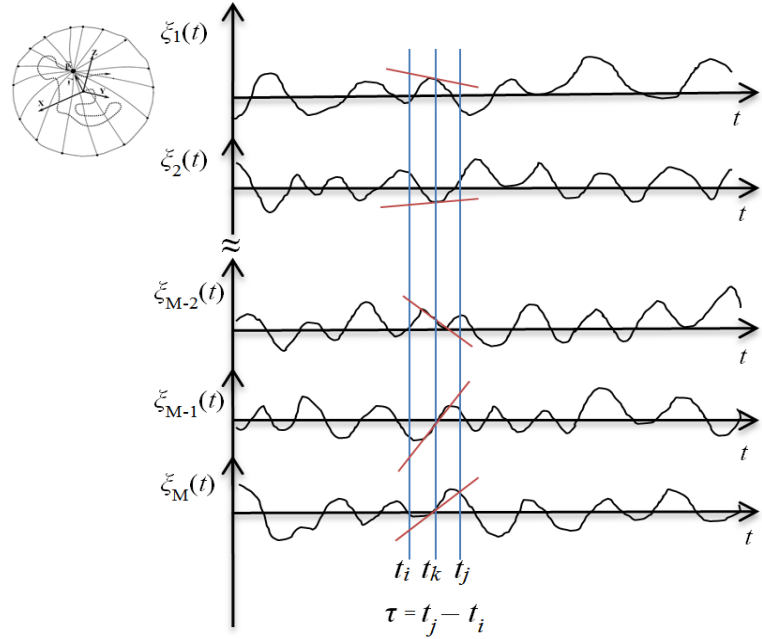

**Fig. A1.1**. The realizations of a differentiable stationary or pseudo-stationary random process $\xi(t)$. These realizations can be interpreted, for example, as time changes in the projection of the location of a wandering particle on the *X* axis (see Figure 1), i.e. $x(t) = \xi(t)$

In General, this process is non-stationary, but we assume that all the averaged characteristics of this process in the section $t_i$ do not significantly differ from its similar averaged characteristics in the section $t_j$. That is, we require that all the moments and central moments of this process in the section $t_i$ are approximately equal to the corresponding moments and central moments in the section $t_j$ when $\tau = t_j - t_i$ tending to zero. For example,

$$\overline{\xi(t_i)} \approx \overline{\xi(t_j)}; \tag{A1.1}$$

$$\overline{\xi^2(t_i)} \approx \overline{\xi^2(t_j)}, \text{ etc.} \tag{A1.2}$$

In other words, the considered random process $\xi(t)$ is either stationary or close to it. However, in each section $t_m$, all the averaged characteristics of such a process remain unchanged. For convenience, we will call such a process "pseudo-stationary random process" (PSRP).

All conclusions about the PSRP, made in this appendix, also apply to the stationary random process (SRP).

There is a known procedure for obtaining the PDF $\rho(\xi'_k)$ of the derivative of a random process $\xi'(t) = d\xi(t)/dt$ with a known two-dimensional PDF of a random stationary process [17, 18]

$$\rho(\xi_i, \xi_j) = \rho(\xi_i, t_i; \xi_j, t_j). \tag{A1.3}$$

In Ex. (A1.3) we make the change of variables

$$\xi_i = \xi_k - \frac{\tau}{2}\xi'_k; \quad \xi_j = \xi_k + \frac{\tau}{2}\xi'_k; \quad t_i = t_k - \frac{\tau}{2}; \quad t_j = t_k + \frac{\tau}{2}, \tag{A1.4}$$

where
$$\tau = t_j - t_i; \quad t_k = \frac{t_j - t_i}{2}, \tag{A1.5}$$

with the Jacobian $[J] = \tau$.

As a result, from the two-dimensional PDF (A1.3) we obtain

$$\rho_2(\xi_k, \xi'_k) = \lim_{\tau \to 0} \tau \, \rho_2\left(\xi_k - \frac{\tau}{2}\xi'_k, \, t_k - \frac{\tau}{2}; \, \xi_k + \frac{\tau}{2}\xi'_k, \, t_k + \frac{\tau}{2}\right). \tag{A1.6}$$

Integrating (A1.6) over $\xi_k$, we find the required PDF $\rho(\xi'_k)$ in the section $t_k$ [17]:

$$\rho(\xi'_k) = \int_{-\infty}^{\infty} \rho(\xi_k, \xi'_k) d\xi_k. \tag{A1.7}$$

Let's now consider the possibility of obtaining the PDF $\rho(\xi'_k)$ for a known one-dimensional PDF $\rho(\xi)$.

To solve this problem, we use the following properties of random processes:

1. A two-dimensional PDF of a random process can be represented as [17,18]

$$\rho(\xi_i, t_i; \xi_j, t_j) = \rho(\xi_i, t_i)\rho(\xi_j, t_j / \xi_i, t_i), \tag{A1.8}$$

where $\rho(\xi_j, t_j/\xi_i, t_i)$ is the conditional PDF.

2. For any PSRP and SRP the approximate identity is valid

$$\rho(\xi_i, t_i) \approx \rho(\xi_j, t_j). \tag{A1.9}$$

3. The conditional PDF of a random process at $\tau = t_i - t_j$ tending to zero becomes in the delta function

$$\lim_{\tau \to 0} \rho(\xi_j, t_j / \xi_i, t_i) = \delta(\xi_j - \xi_i). \tag{A1.10}$$

Using the above properties, we prepare a random process in the interval $[t_i = t_k - \tau/2; t_j = t_k + \tau/2]$ as $\tau \to 0$, based on the following procedure.

The PDF $\rho(\xi_i) = \rho(\xi_i, t_i)$ in the section $t_i$ and the PDF $\rho(\xi_j) = \rho(\xi_j, t_j)$ in the section $t_j$ can always be represented as a product of two functions

$$\rho(\xi_i) = \varphi(\xi_i)\varphi(\xi_i) = \varphi^2(\xi_i), \tag{A1.11}$$

$$\rho(\xi_j) = \varphi(\xi_j)\varphi(\xi_j) = \varphi^2(\xi_j),$$

where $\varphi(\xi_i)$ is the probability amplitude (PA) of the random variable $\xi_i$ in the section $t_i$; $\varphi(\xi_j)$ is a PA of a random variable $\xi_j$ in the section $t_j$.

For PSRP, the approximate expression is valid

$$\varphi(\xi_i) \approx \varphi(\xi_j), \tag{A1.12}$$

which can be verified by taking the square root of both parts (A1.9).

For SRP, the approximate relation (A1.12) becomes the equality

$$\varphi(\xi_i) = \varphi(\xi_j). \tag{A1.12a}$$

Note that the approximate Ex. (A.1.12) at $\tau \to 0$ for the majority of non-stationary random processes (including for PSRP) also turns into the equality

$$\varphi(\xi_i, t_i) = \lim_{\tau \to 0} \varphi(\xi_j, t_j = t_i + \tau). \tag{A1.13}$$

When the condition (A.1.12) [or (A.1.12a)] is satisfied, Ex. (A1.8) can be represented in the following form

$$\rho(\xi_i, \xi_j) \approx \varphi(\xi_i)\rho(\xi_j / \xi_i)\varphi(\xi_j), \tag{A1.14}$$

where $\rho(\xi_j / \xi_i)$ is the conditional PDF.

Let's write (A.1.14) in expanded form

$$\rho\left[\xi_i, t_i = t_k - \frac{\tau}{2}; \xi_j, t_j = t_k + \frac{\tau}{2}\right] \approx$$
$$\approx \left[\xi_i, t_i = t_k - \frac{\tau}{2}\right]\rho\left[\xi_j, t_j = t_k + \frac{\tau}{2} / \xi_i, t_i = t_k - \frac{\tau}{2}\right]\varphi\left[\xi_j, t_j = t_k + \frac{\tau}{2}\right]. \quad (A1.15)$$

Let $\tau$ tend to zero in (A1.15), so that the given time interval contracts uniformly on the left and right at the middle moment of time $t_k = (t_j+t_i)/2$. In this case, taking into account (A1.10), from (A1.14), we obtain the exact equality

$$\lim_{\tau \to 0}\rho(\xi_i, \xi_j) = \lim_{\tau \to 0}\{\varphi(\xi_i)\rho(\xi_j/\xi_i)\varphi(\xi_j)\} = \varphi(\xi_{ik})\delta(\xi_{jk} - \xi_{ik})\varphi(\xi_{jk}), \quad (A1.16)$$

where $\xi_{ik}$ is the result of the tendency of the random variable $\xi(t_i)$ to the random variable $\xi(t_k)$ on the left; $\xi_{jk}$ is the result of the tendency of the random variable $\xi(t_j)$ to the random variable $\xi(t_k)$ from the right.

Integrating both sides of Ex. (A1.16) over $\xi_{ik}$ and $\xi_{jk}$, we obtain

$$\int_{-\infty}^{\infty}\int_{-\infty}^{\infty}\varphi(\xi_{ik})\delta(\xi_{jk} - \xi_{ik})\varphi(\xi_{jk})d\xi_{ik}d\xi_{jk} = 1. \quad (A1.17)$$

In (A1.17), the properties of the $\delta$-function are taken into account.

Let's set the specific form of the $\delta$-function. To do this, consider a random Markov process for which the diffusion equation (a particular case of the Fokker-Planck - Kolmogorov equation) is valid

$$\frac{\partial \rho(\xi_j/\xi_i)}{\partial t} = B\frac{\partial^2 \rho(\xi_j/\xi_i)}{\partial \xi^2}, \quad (A1.18)$$

where $B$ is the diffusion coefficient.

One of the solutions of this differential equation, as is well known, has the form

$$\rho(\xi_j, t_j / \xi_i, t_i) = \frac{1}{2\pi}\int_{-\infty}^{\infty}\exp\{iq(\xi_j - \xi_i) - q^2 B(t_j - t_i)\}dq, \quad (A1.19)$$

where $q$ is the generalized frequency.

For $\tau = t_j - t_i \to 0$ from (A.1.19) we obtain one of the definitions of the $\delta$-function

$$\lim_{\tau \to 0} \rho(\xi_j / \xi_i) = \frac{1}{2\pi} \int_{-\infty}^{\infty} \exp\{iq(\xi_{jk} - \xi_{ik})\} dq = \delta(\xi_j - \xi_i). \quad (A1.20)$$

*This result was obtained for the case as τ → 0. Therefore, the δ-function (A1.20) can correspond not only to a Markov random process, but also to many other stationary and non-stationary random processes. In other words, one could immediately assume that the δ-function for the PSRP has the form (A.1.20) without referring to the equation (A.1.18).*

Let's substitute δ-function (A.1.20) into Ex. (A.1.17)

$$\int_{-\infty}^{\infty}\int_{-\infty}^{\infty} \varphi(\xi_{ik}) \left[\frac{1}{2\pi}\int_{-\infty}^{\infty} \exp\{iq(\xi_{jk}-\xi_{ik})\}dq\right] \varphi(\xi_{jk}) d\xi_{ik} d\xi_{jk} = 1, \quad (A1.21)$$

and change the order of integration in (A1.21)

$$\int_{-\infty}^{\infty}\left[\frac{1}{\sqrt{2\pi}}\int_{-\infty}^{\infty} \varphi(\xi_{ik})\exp\{-iq\xi_{ik}\}d\xi_{ik} \frac{1}{\sqrt{2\pi}}\int_{-\infty}^{\infty} \varphi(\xi_{jk})\exp\{iq\xi_{jk}\}d\xi_{jk}\right] dq = 1. \quad (A1.22)$$

Let's take into account that, according to (A1.13), for the SRP and PSRP the condition $\varphi(\xi_{ik}) = \varphi(\xi_{jk})$ is fulfilled. Therefore, Ex. (A.1.22) can be represented as

$$\int_{-\infty}^{\infty} \varphi(q)\varphi^*(q) dq = 1, \quad (A1.23)$$

where
$$\varphi(q) = \frac{1}{\sqrt{2\pi}} \int_{-\infty}^{\infty} \varphi(\xi_k) \exp\{-iq\xi_k\} d\xi_k, \quad (A1.24)$$

$$\varphi^*(q) = \frac{1}{\sqrt{2\pi}} \int_{-\infty}^{\infty} \varphi(\xi_k) \exp\{iq\xi_k\} d\xi_k. \quad (A1.25)$$

The integrand $\varphi(q)\varphi^*(q)$ in (A1.23) meets all the requirements of the PDF $\rho(q)$ of the random variable $q$:

$$\rho(q) = \varphi(q)\varphi^*(q) = |\varphi(q)|^2. \quad (A1.26)$$

Let's clarify the physical meaning of $q$.

The features of the considered random process impose the following restrictions on the generalized frequency $q$:

1) the random variable $q$ should characterize the random process $\xi(t)$ in the section $t_k$ (see Figure A1.1), i.e. in the interval $\tau = t_j - t_i$ tending to 0;

2) the variable $q$ must belong to the set of real numbers, that is, take any value from the range $]-\infty, \infty[$.

These requirements are satisfied by the following random values associated with the PSRP (or SRP) in the time interval $\tau$:

$$\xi_i' = \frac{\partial \xi_k}{\partial t}, \quad \xi_i'' = \frac{\partial^2 \xi_k}{\partial t}, \quad \ldots, \quad \xi_i^{(n)} = \frac{\partial^n \xi_i}{\partial t^n}. \tag{A1.27}$$

To clarify which of these values is associated with the generalized frequency $q$, consider one implementation of the process under study (see Figure A1.1). The function $\xi(t)$ in the interval at $\tau < \tau_{cor}$ [where $\tau_{cor}$ is the autocorrelation interval of the random process $\xi(t)$] can be expanded in the Maclaurin series

$$\xi(t_j) = \xi(t_i) + \xi'(t_i)\tau + \frac{\xi''(t_i)}{2}\tau^2 + \ldots + \frac{\xi^{(n)}(t_i)}{n!}\tau^n + \ldots \tag{A1.28}$$

The Ex. (A1.28) is presented in the following form

$$\frac{\xi_j - \xi_i}{\tau} = \xi_i' + \frac{\xi_i''}{2!}\tau + \ldots + \frac{\xi_i^{(n)}\tau^{n-1}}{n!} + \ldots \tag{A1.29}$$

where $\xi(t_i) = \xi_i$, $\xi(t_j) = \xi_j$, and we tend $\tau$ to zero.

In this case (A1.29) is reduced to the expression

$$\lim_{\tau \to 0} \frac{\xi_j - \xi_i}{\tau} = \xi_k', \tag{A1.30}$$

where $\xi_k = \xi(t_k)$ (see Figure A1.1).

Therefore, it remains to assume that the generalized frequency $q$ in Ex.s (A1.23) – (A1.26) is linearly related only to $\xi'_k$, i.e.

$$q = \frac{\xi_k'}{\eta}, \tag{A1.31}$$

where $\eta$ is the scale parameter.

The Ex. (A1.31) can be obtained in another way.

Each exponential, for example, from the integral (A1.24), corresponds to a harmonic function with frequency $q$

$$\exp\{-iq\xi(t)\} \to \xi_k(t) = A\sin(qt), \quad (A1.32)$$

this is one of the harmonic components of the random process $\xi(t)$.

Each frequency $q$, in turn, corresponds to the tangent of the angle of inclination of the tangent line to the harmonic function with a given frequency (see Figure A1.2), that is, $q \sim tg\alpha = \xi'(t)$.

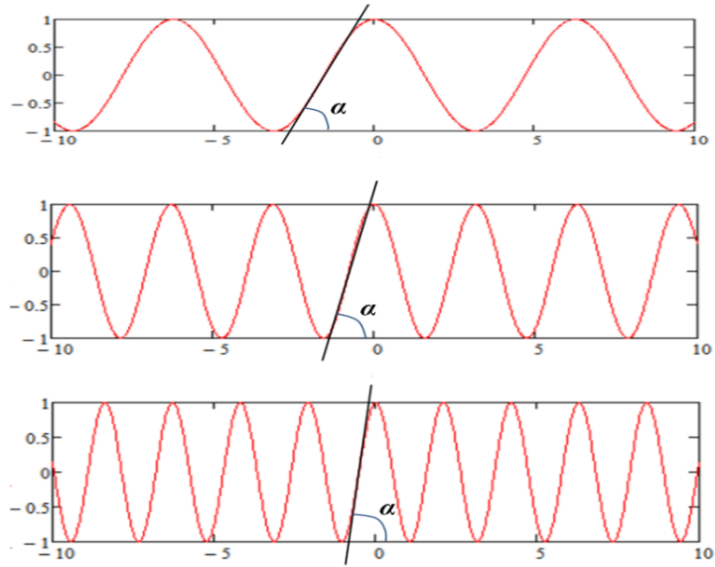

**Fig. A1.2.** The greater the frequency $q$ of the harmonic function, the greater the angle $\alpha$ between the tangent to this function and the $t$ axis

Differentiating Ex. (A1.32), we obtain $\xi'_k(t) = qA\cos(qt)$, whence it follows

$$q = \lim_{t \to 0} \frac{\xi'_k}{A\cos(qt)} = \frac{\xi'_k}{A}. \quad (A1.33)$$

For $A = \eta$ Ex.s (A1.31) and (A1.33) coincide.

Substituting (A1.31) into (A1.23) – (A1.26), we obtain the following required procedure for obtaining the PDF $\rho(\xi',t)$ of a pseudo-stationary random process (PSRP) or stationary random process (SRP) $\xi(t)$ in any section $t_k$ for a known one-dimensional PDF $\rho(\xi,t)$ in the same section:

1. A given one-dimensional PDF $\rho(\xi,t)$ is represented as a product of two probability amplitudes (PA) $\varphi(\xi)$:

$$\rho(\xi,t) = \varphi(\xi,t)\varphi(\xi,t). \qquad (A1.34)$$

2. Two Fourier transforms are performed

$$w(\xi',t) = \frac{1}{\sqrt{2\pi\eta}} \int_{-\infty}^{\infty} \varphi(\xi,t)\exp\{i\xi'\xi/\eta\}d\xi, \qquad (A1.35)$$

$$w*(\xi',t) = \frac{1}{\sqrt{2\pi\eta}} \int_{-\infty}^{\infty} \varphi(\xi,t)\exp\{-i\xi'\xi/\eta\}d\xi. \qquad (A1.36)$$

3. Finally, for an any section of the PSRP (or SRP), we obtain the required PDF of its derivative

$$\rho(\xi',t) = w(\xi',t)w^*(\xi',t) = |w(\xi',t)|^2. \qquad (A1.37)$$

Once again, we note that the procedure (A1.34) – (A1.37) can be applied to any stationary and pseudo-stationary random processes {i.e., random processes with a slowly varying PDF $\rho(\xi,t)$}, for which, as $\tau \to 0$, the $\delta$-function takes the form (A1.20).

To clarify the physical meaning of the scale parameter $\eta$, consider a stationary random process $\xi(t)$ with a Gaussian distribution of the random variable $\xi$

$$\rho(\xi) = \frac{1}{\sqrt{2\pi\sigma_\xi^2}} \exp\{-(\xi - a_\xi)^2/2\sigma_\xi^2\}, \qquad (A1.38)$$

where $\sigma_\xi^2$ and $a_\xi$ are the variance and mathematical expectation of the process.

Performing the sequence of operations (A1.34) – (A1.37) with the PDF (A1.38), we obtain the PDF of the derivative of this random process

$$\rho(\xi') = \frac{1}{\sqrt{2\pi[\eta/2\sigma_\xi]^2}} \exp\left\{-\frac{\xi'^2}{2[\eta/2\sigma_\xi]^2}\right\}. \qquad (A1.39)$$

On the other hand, using the well-known procedure (A1.4) – (A1.7) for a similar case, we obtain [18]

$$\rho(\xi') = \frac{1}{\sqrt{2\pi\sigma_{\xi'}^2}} \exp\{-\xi'^2/2\sigma_{\xi'}^2\}, \quad (A1.40)$$

where
$$\sigma_{\xi'} = \sigma_\xi/\tau_{\xi\,cor},$$

here $\tau_{\xi\,cor}$ is the autocorrelation interval of the initial random process $\xi(t)$.

Comparing the PDF (A1.39) and (A1.40), we find that

$$\eta = \frac{2\sigma_\xi^2}{\tau_{\xi\,cor}}. \quad (A1.41)$$

The Ex. (A1.41) was obtained for a Gaussian random process, but the standard deviation $\sigma_\xi$ and the autocorrelation interval $\tau_{\xi\,cor}$ are the main characteristics of any SRP or PSRP. All other moments and central moments in the case of a non-Gaussian distribution of the random variable $\xi(t)$ will make an insignificant contribution to the Ex. (A1.41). Therefore, it can be argued with a high degree of reliability that Ex. (A1.41) is applicable for a large class of stationary and pseudo-stationary random processes.

In quantum mechanics for the transition from the coordinate representation of the wave function of a pico-particle to its momentum representation, there is the procedure

$$\varphi(p_x) = \frac{1}{\sqrt{2\pi\hbar}} \int_{-\infty}^{\infty} \psi(x)\exp\{ip_x x/\hbar\}dx = \frac{1}{\sqrt{2\pi\hbar}} \int_{-\infty}^{\infty} \psi(x)\exp\{imx'x/\hbar\}dx, \quad (A1.42)$$

$$\varphi*(p_x) = \frac{1}{\sqrt{2\pi\hbar}} \int_{-\infty}^{\infty} \psi(x)\exp\{-ip_x x/\hbar\}dx = \frac{1}{\sqrt{2\pi\hbar}} \int_{-\infty}^{\infty} \psi(x)\exp\{-imx'x/\hbar\}dx, \quad (A1.43)$$

where $\hbar = 1.055\times10^{-34}$ J/Hz is the reduced Planck's constant, and it is also taken into account that the $x$-component of the particle momentum $p_x$ is related with its speed $v_x$ (i.e., time derivative)

$$p_x = mv_x = m\frac{dx}{dt} = mx'. \quad (A1.45)$$

In the case when

$$\eta_x = \frac{2\sigma_x^2}{\tau_{xcor}} = \frac{\hbar}{m} \quad \text{with dimension } (m^2/s), \tag{A1.46}$$

procedures (A1.34) – (A1.37) and (A1.42) – (A1.43) completely coincide.

From Ex. (A1.46) it follows that Planck's constant can be expressed through the main averaged parameters $\sigma_x$ and $\tau_{x\,cor}$ of a stationary (or pseudo-stationary) random process, which involves a randomly wandering pico-particle (for example, an electron).

At the same time, the field of application of the procedure (A1.42) – (A1.43) is limited by the smallness of the reduced Planck's constant $\hbar$. While the procedure (A1.34) – (A1.37) can be applied for random stationary and pseudo-stationary processes of any scale. Such random processes include chaotic oscillations of the center of mass of the nucleus of a biological cell, chaotic movements of the tip of a tree branch, chaotic change in the position of the center of mass of the planet's nucleus, etc.

Let's note the following intermediate results:

1]. For a stationary and pseudo-stationary random process $\xi(t) = x(t)$, the following procedure for obtaining the PDF $\rho(x')$ derivative of this process can be applied.

A given one-dimensional PDF $\rho(x)$ of a stationary process [or a slowly varying PDF $\rho(x,t)$ of a pseudo-stationary process] is represented as a product of two PA $\varphi(x)$ or $\varphi(x,t)$:

$$\rho(x) = \varphi(x)\varphi(x) \quad \text{or} \quad \rho(x,t) = \varphi(x,t)\varphi(x,t). \tag{A1.47}$$

*a*) For a stationary random process (SRP), two Fourier transforms are performed

$$\varphi(x') = \frac{1}{\sqrt{2\pi\eta_x}} \int_{-\infty}^{\infty} \psi(x) \exp\{ix'x/\eta_x\} dx = \frac{1}{\sqrt{2\pi\eta_x}} \int_{-\infty}^{\infty} \psi(x) \exp\{iv_x x/\eta_x\} dx, \tag{A1.48}$$

$$\varphi*(x') = \frac{1}{\sqrt{2\pi\eta_x}} \int_{-\infty}^{\infty} \psi(x) \exp\{-ix'x/\eta_x\} dx = \frac{1}{\sqrt{2\pi\eta_x}} \int_{-\infty}^{\infty} \psi(x) \exp\{-iv_x x/\eta_x\} dx, \tag{A1.49}$$

and the desired PDF of the derivative of this process is determined

$$\rho(x') = \varphi(x')\varphi^*(x') = |\varphi(x')|^2, \quad (A1.50)$$

or
$$\rho(v_x) = \varphi(v_x)\varphi^*(v_x) = |\varphi(v_x)|^2, \quad (A1.51)$$

where
$$\eta_x = \frac{2\sigma_x^2}{\tau_{x\,\kappa op}} = \frac{\hbar}{m} \quad (A1.52)$$

$\sigma_x$ is the standard deviation of the initial stationary random process $x(t)$;

$\tau_{x\,cor}$ is the autocorrelation interval of this process.

In § 2.6 of the article [11, arXiv:2007.13527], the procedure (A1.47) – (A1.51) is applied to obtain the PDF $\rho(x')$ of the derivative of stationary random processes with distribution laws: Gaussian, uniform, Laplace, Cauchy and sinusoidal.

b) For a pseudo-stationary random process (PSRP), two Fourier transforms are performed

$$\varphi(x',t) = \frac{1}{\sqrt{2\pi\eta_x(t)}} \int_{-\infty}^{\infty} \psi(x,t)\exp\{ix'x/\eta_x(t)\}dx = \frac{1}{\sqrt{2\pi\eta_x(t)}} \int_{-\infty}^{\infty} \psi(x,t)\exp\{iv_x x/\eta_x(t)\}dx, \quad (A1.53)$$

$$\varphi^*(x',t) = \frac{1}{\sqrt{2\pi\eta_x(t)}} \int_{-\infty}^{\infty} \psi(x,t)\exp\{-ix'x/\eta_x(t)\}dx = \frac{1}{\sqrt{2\pi\eta_x(t)}} \int_{-\infty}^{\infty} \psi(x,t)\exp\{-iv_x x/\eta_x(t)\}dx, \quad (A1.54)$$

and the required PDF of the derivative of this process is determined at each time moment $t$

$$\rho(x',t) = \varphi(x',t)\varphi^*(x',t) = |\varphi(x',t)|^2, \quad (A1.55)$$

or
$$\rho(v_x,t) = \varphi(v_x,t)\varphi^*(v_x,t) = |\varphi(v_x,t)|^2, \quad (A1.56)$$

where
$$\eta_x(t) = \frac{2\sigma_x^2(t)}{\tau_{x\,cor}(t)} = \frac{\hbar}{m(t)}; \quad (A1.57)$$

$\sigma_x(t)$ is the standard deviation of the initial pseudo-stationary random process $x(t)$ from its mean value at time $t$;

$\tau_{x\,cor}(t)$ is the autocorrelation interval of this process at time $t$.

2]. The procedure (A1.47) – (A1.52) up to the proportionality coefficient $\eta$ coincides with the quantum-mechanical procedure (A1.42) – (A1.43) of transition from the coordinate representation to the impulse one. But the quantum - mechanical procedure (A1.42) – (A1.43) was obtained using a rather unobvious (exotic) hypothesis about the possible existence of de Broglie's waves of matter (which were never discovered). While the procedure (A1.47) – (A1.52) is obtained on the basis of a detailed analysis of a differentiable random process with the only assumption (which may be questioned) that the $\delta$-function has the form (A1.20). In this regard, it is interesting to analyze which procedures for the transition from PDF $\rho(x)$ to PDF $\rho(x')$ can lead to other types of $\delta$-function?

Also, there is no need to use Louis de Broglie's hypothesis of the existence of matter waves to describe the diffraction of particles by a crystal. In the article [11, arXiv:2007.13527 ] it is shown that, based on the laws of geometric optics and the theory of probability, a formula was obtained for calculating the volumetric scattering diagrams of particles on a multilayer periodic surface of a crystal.

3]. In the case of studying the chaotic behavior of pico-particles, the ratio $\hbar/m$ can be expressed through the main characteristics of the investigated random process (A1.46). In the author's opinion, this is a very important result, since it is practically impossible to estimate the real mass of a mobile elementary particle. Let's recall that in physical reference books only the rest masses of elementary particles are given, which are determined indirectly on the basis of complex experiments. Whereas it is much easier to obtain an estimate of the standard deviation $\sigma_x$ and the autocorrelation interval $\tau_{x\,cor}$ of a randomly wandering particle. It is also important that the reduced Planck constant $\hbar$ loses its fundamental character and turns out to be the dimensional coefficient of proportionality between the particle mass and the ratio of the averaged characteristics of the random process.

Appendix 2

## A2 Coordinate representation of a characteristics of the chaotically wandering particle (ChWP)

### A2.1 Coordinate representation of an average speed of the ChWP

For stationary and pseudo-stationary random processes (see Appendix1), we prove the validity of equalities

$$\overline{x'^n} = \overline{v_x^n} = \int_{-\infty}^{+\infty} \rho(x') x'^n dx'_x = \int_{-\infty}^{+\infty} \rho(v_x) v_x^n dv_x = \int_{-\infty}^{+\infty} \phi(v_x) v_x^n \phi(v_x) dv_x =$$
$$= \int_{-\infty}^{+\infty} \psi(x) \left( \pm i\eta_x \frac{\partial}{\partial x} \right)^n \psi(x) dx, \qquad (A2.1)$$

and

$$\overline{x'^n}(t) = \overline{v_x^n}(t) = \int_{-\infty}^{+\infty} \rho(x', t) x'^n dx'_x = \int_{-\infty}^{+\infty} \rho(v_x, t) v_x^n dv_x =$$
$$= \int_{-\infty}^{+\infty} \phi(v_x, t) v_x^n \phi(v_x, t) dv_x = \int_{-\infty}^{+\infty} \psi(x, t) \left( \pm i\eta_x \frac{\partial}{\partial x} \right)^n \psi(x, t) dx, \qquad (A2.2)$$

where $n$ is an integer, positive degree; $\eta_x$ is the scale parameter (A1.52).

Experts in the field of QM are well aware of the proof of a similar expression

$$\overline{p_x^n} = \int_{-\infty}^{+\infty} \rho(p_x) p_x^n dp_x = \int_{-\infty}^{+\infty} \psi(p_x) p_x^n \psi(p_x) dp_x = \int_{-\infty}^{+\infty} \psi(x) \left( -i\hbar \frac{\partial}{\partial x} \right)^n \psi(x) dx,$$

see, for example, [20]. However, in view of the importance of this proof for this article, we present it in a slightly modified form, as applied to the features of mass-independent stochastic quantum mechanics (MSQM).

Let's use the Fourier transforms (A1.48) and (A1.49)

$$\phi(v_x) = = \int_{-\infty}^{+\infty} \psi(x) \frac{e^{i\frac{v_x x}{\eta_x}}}{(2\pi\eta_x)^{1/2}} dx = \int_{-\infty}^{+\infty} \psi(x) \frac{e^{i\frac{p_x x}{\hbar}}}{(2\pi\hbar)^{1/2}} dx, \qquad (A2.3)$$

$$\phi^*(v_x) = \int_{-\infty}^{+\infty} \psi(x) \frac{e^{-i\frac{v_x x}{\eta_x}}}{(2\pi\eta_x)^{\frac{1}{2}}} dx = \int_{-\infty}^{+\infty} \psi(x) \frac{e^{-i\frac{p_x x}{\hbar}}}{(2\pi\hbar)^{\frac{1}{2}}} dx. \qquad (A2.4)$$

Substitute integrals (A2.3) and (A2.4) into the third part of Eq. (A2.1)

$$\overline{v_x^n} = \int_{-\infty}^{+\infty} \left[ \int_{-\infty}^{+\infty} \psi(x_i) \frac{e^{-i\frac{v_x x_i}{\eta_x}}}{(2\pi\eta_x)^{1/2}} dx_i v_x^n \int_{-\infty}^{+\infty} \psi(x_j) \frac{e^{i\frac{v_x x_j}{\eta_x}}}{(2\pi\eta_x)^{1/2}} dx_j \right] dv_x. \qquad (A2.5)$$

---

It is easy to verify by direct verification that

$$v_x^n e^{i\frac{v_x x_j}{\eta_x}} = \left(-i\eta_x \frac{\partial}{\partial x_j}\right)^n e^{i\frac{v_x x_j}{\eta_x}}, \quad \text{или} \quad p_x^n e^{i\frac{p_x x_j}{\hbar}} = \left(-i\hbar \frac{\partial}{\partial x_j}\right)^n e^{i\frac{p_x x_j}{\hbar}}, \quad (A2.6)$$

$$v_x^n e^{-i\frac{v_x x_i}{\eta_x}} = \left(i\eta_x \frac{\partial}{\partial x_i}\right)^n e^{-i\frac{v_x x_i}{\eta_x}}, \quad \text{или} \quad p_x^n e^{i\frac{p_x x_i}{\hbar}} = \left(i\hbar \frac{\partial}{\partial x_i}\right)^n e^{-i\frac{p_x x_i}{\hbar}}. \quad (A2.6a)$$

Let's rewrite (A2.5) taking into account (A2.6)

$$\overline{v_x^n} = \frac{1}{2\pi\eta_x} \int_{-\infty}^{+\infty} \left[ \int_{-\infty}^{+\infty} \psi(x_i) e^{-i\frac{v_x x_i}{\eta_x}} dx_i \int_{-\infty}^{+\infty} \psi(x_j) \left(-i\eta_x \frac{\partial}{\partial x_j}\right)^n e^{i\frac{v_x x_j}{\eta_x}} dx_j \right] dv_x.$$

or taking into account (A2.6a)                                                    (A2.7a)

$$\overline{v_x^n} = \frac{1}{2\pi\eta_x} \int_{-\infty}^{+\infty} \left[ \int_{-\infty}^{+\infty} \psi(x_i) \left(i\eta_x \frac{\partial}{\partial x_i}\right)^n e^{-i\frac{v_x x_i}{\eta_x}} dx_i \int_{-\infty}^{+\infty} \psi(x_j) e^{i\frac{v_x x_j}{\eta_x}} dx_j \right] dv_x.$$

We integrate the second integral in the integrand (A2.7) *n* times by parts, and we will assume that *ψ*(*x*) and its derivatives vanish at the integration boundaries *x* = ± ∞. Performing these actions with expression (A2.7), we get [20]

$$\overline{v_x^n} = \frac{1}{2\pi\eta_x} \int_{-\infty}^{+\infty} \left[ \int_{-\infty}^{+\infty} \psi(x_i) e^{-i\frac{v_x x_i}{\eta_x}} dx_i \int_{-\infty}^{+\infty} e^{i\frac{v_x x_j}{\eta_x}} \left(-i\eta_x \frac{\partial}{\partial x_j}\right)^n \psi(x_j) dx_j \right] dv_x,$$

or                                                                                                           (П2.8)

$$\overline{v_x^n} = \frac{1}{2\pi\eta_x} \int_{-\infty}^{+\infty} \left[ \int_{-\infty}^{+\infty} dx_i \int_{-\infty}^{+\infty} \psi(x_i) e^{i\frac{v_x(x_j - x_i)}{\eta_x}} \left(-i\eta_x \frac{\partial}{\partial x_j}\right)^n \psi(x_j) dx_j \right] dv_x.$$

(П2.9)

Similarly, we integrate the second integral in the integrand (A2.7*a*) *n* times by parts, and we will assume that *ψ*(*x*) and its derivatives vanish at the integration boundaries *x* = ± ∞. Performing these actions with expression (A2.7), we get

$$\overline{v_x^n} = \frac{1}{2\pi\eta_x} \int_{-\infty}^{+\infty} \left[ \int_{-\infty}^{+\infty} e^{-i\frac{v_x x_i}{\eta_x}} \left(i\eta_x \frac{\partial}{\partial x_i}\right)^n \psi(x_i) dx_i \int_{-\infty}^{+\infty} e^{i\frac{v_x x_j}{\eta_x}} \psi(x_j) dx_j \right] dv_x,$$

or                                       (П2.8*a*)

$$\overline{v_x^n} = \frac{1}{2\pi\eta_x} \int_{-\infty}^{+\infty} \left[ \int_{-\infty}^{+\infty} \psi(x_i) \left(i\eta_x \frac{\partial}{\partial x_i}\right)^n \psi(x_j) e^{i\frac{v_x(x_j - x_i)}{\eta_x}} dx_i \int_{-\infty}^{+\infty} dx_j \right] dv_x.$$

(П2.9*a*)

Let's change the order of integration in (A2.9) and (A2.9*a*), i.e. first we will integrate over $v_x$

$$\overline{v_x^n} = \int_{-\infty}^{+\infty} dx_i \int_{-\infty}^{+\infty} dx_j \psi(x_i) \left(-i\hbar \frac{\partial}{\partial x_j}\right)^n \psi(x_j) \frac{1}{2\pi\eta_x} \int_{-\infty}^{+\infty} e^{i\frac{v_x(x_j - x_i)}{\eta_x}} dv_x,$$

$$\overline{v_x^n} = \int_{-\infty}^{+\infty} dx_j \int_{-\infty}^{+\infty} dx_i \psi(x_j) \left(i\hbar \frac{\partial}{\partial x_i}\right)^n \psi(x_i) \frac{1}{2\pi\eta_x} \int_{-\infty}^{+\infty} e^{i\frac{v_x(x_j-x_i)}{\eta_x}} dv_x.$$

There is a delta function in that expressions

$$\delta(x_j - x_i) = \frac{1}{2\pi\eta_x} \int_{-\infty}^{+\infty} e^{i\frac{v_x(x_j-x_i)}{\eta_x}} dv_x \quad \text{type} \quad \delta(x_j - x_i) = \frac{1}{2\pi} \int_{-\infty}^{+\infty} e^{iq(x_j-x_i)} dq. \quad (A2.10)$$

Therefore, we represent it in the form

$$\overline{v_x^n} = \int_{-\infty}^{+\infty} dx_i \int_{-\infty}^{+\infty} \psi(x_i) \left(-i\eta_x \frac{\partial}{\partial x_j}\right)^n \psi(x_j) \delta(x_j - x_i) dx_j. \quad (A2.11)$$

$$\overline{v_x^n} = \int_{-\infty}^{+\infty} dx_i \int_{-\infty}^{+\infty} \psi(x_i) \left(i\eta_x \frac{\partial}{\partial x_j}\right)^n \psi(x_j) \delta(x_j - x_i) dx_j. \quad (A2.11a)$$

Using the properties of the $\delta$-function, we finally write

$$\overline{v_x^n} = \int_{-\infty}^{+\infty} \psi(x) \left(\mp i\eta_x \frac{\partial}{\partial x}\right)^n \psi(x) dx, \quad (A2.12)$$

$$\eta_x = \frac{2\sigma_x^2}{\tau_{xкор}} = const. \quad (A2.12a)$$

thus, Ex. (A2.1) is proved for the case of a stationary random process (SSP).

For a pseudo-stationary random process (PSRP), Ex. (A2.2) is proved similarly. Performing operations similar to (A2.5) – (A2.15) using transformations (A1.53) and (A1.54), we obtain

$$\overline{v_x^n}(t) = \int_{-\infty}^{+\infty} \psi(x,t) \left(\mp i\eta_x \frac{\partial}{\partial x}\right)^n \psi(x,t) dx. \quad (A2.13)$$

where $\quad \eta_x = \frac{2\sigma_x^2}{\tau_{xкор}} \approx \eta_x(t) \approx \frac{2\sigma_x^2(t)}{\tau_{xкор}(t)} \approx const. \quad (A2.13a)$

In the general case, the scale parameter (A1.57) can change with time $\eta_x(t)$. However, in many non-stationary stochastic systems, it remains unchanged, since the variance $\sigma_x^2(t)$ and the autocorrelation radius $\tau_{xcor}(t)$ of the pseudo-stationary random process (PSRP) change simultaneously and proportionally with respect to each other. For example, a situation is possible when the variance of the pseudo-stationary random process (PSRP) changes with time according to the law $\sigma_x^2(t) = \sigma_x^2 \times (t - t_0)$, and its autocorrelation coefficient changes according to the same law $\tau_{x\,cor}(t) = \tau_{x\,cor} \times (t - t_0)$, then

$$\eta_x(t) = \frac{2\sigma_x^2(t)}{\tau_{x\kappa op}(t)} \approx \frac{2\sigma_x^2 \times (t-t_0)}{\tau_{x\kappa op} \times (t-t_0)} \approx \eta_x = \frac{2\sigma_x^2}{\tau_{x\kappa op}} = const. \quad (A2.14)$$

The invariable ratio of the main averaged characteristics of the investigated random process will be called: "*The law of proportional constancy of the scale parameter of the stochastic system.*"

## A2.2 Coordinate representation of the globally averaged change in the mechanical energiality of the ChWP

Let's consider the case when a change in the probability amplitude $\psi(x,t)$ is associated with a change in the kinetic energiality of the ChWP.

Let's return to the consideration of the conditional PDF (A1.19)

$$\rho(x_j, t_j / x_i, t_i) = \frac{1}{2\pi} \int_{-\infty}^{\infty} \exp\{iq(x_j - x_i) - q^2 B(t_j - t_i)\} dq, \quad (A2.15)$$

where, according to (A1.31) and under condition (A2.14)

$$q = \frac{x'}{\eta_x} = \frac{v_x}{\eta_x}. \quad (A2.16)$$

If $\Delta x = x_j - x_i \to 0$, that from Ex. (A2.16) we obtain

$$\lim_{\Delta x \to 0} \rho(x_j, t_j / x_i, t_i) = \frac{1}{2\pi\eta_x} \int_{-\infty}^{\infty} \exp\{-\frac{v_x^2}{\eta_x^2} B(t_j - t_i)\} dv_x. \quad (A2.17)$$

Let's take into account that

$$\frac{v_x^2}{2} = k_x, \quad (A2.18)$$

where $k_x$ is the x-kinetic energiality of the particle, according to the terminology of mass-independent physics (14) – (16), i.e. physics freed from the heuristic concept of "mass".

We also take into account that for some stochastic processes without friction, it should be assumed that the diffusion coefficient $B$ is a complex quantity

$$B = iD. \tag{A2.19}$$

*** The self-diffusion coefficient B is complex because the ChWP under consideration alternately diffuses in the chaotically fluctuating medium surrounding it, either in the forward direction or in the opposite direction in the vicinity of the conditional center of the stochastic system (see Figure 1). This vibrational behavior of the ChWP is due to the fact that it is alone and there is no pressure on it from the side of particles similar to it. Self-diffusion of ChWP occurs only due to the averaged excess (or lack) of chaotic exchange of kinetic energiality (i.e., the intensity of movement) between ChWP and the environment.*

*In the general case, $B = e^{-(\mu_0 - i\mu)} = De^{i\mu} = = D[\cos(\mu) + i\sin(\mu)]$, where $\mu_0$ and $\mu$ are the parameters of complex self-diffusion, $D = e^{-\mu_0}$. In the particular case, for $\mu = \pi/2$, it turns out that $B = 0 + iD$. For $D = \eta_x$ the considered stochastic system turns out to be self-consistent.*

Using expressions (A2.18) and (A2.19), we write equation (A2.17) in the following form

$$\lim_{\Delta x \to 0} \rho(x_j, t_j / x_i, t_i) = \frac{1}{2\pi\eta_x} \int_{-\infty}^{\infty} \exp\{-i \frac{v_x^2}{2} \frac{2D}{\eta_x^2} (t_j - t_i)\} dv_x. \tag{A2.20}$$

Substitute (A2.20) into an expression like Ex. (A.1.17)  (A2.21)

$$\int_{-\infty}^{\infty} \left[ \int_{-\infty}^{\infty} \int_{-\infty}^{\infty} \psi(x, t_i) \frac{1}{2\pi\eta_x} \exp\{-i \frac{v_x^2}{2} \frac{2D}{\eta_x^2} (t_j - t_i)\} \psi(x, t_j) dt_i dt_j \right] dv_x = 1.$$

and change the order of integration  (A2.22)

$$\int_{-\infty}^{\infty} \left[ \int_{-\infty}^{\infty} \int_{-\infty}^{\infty} \psi(x, t_i) \frac{1}{2\pi\eta_x} \exp\{-i \frac{v_x^2}{2} \frac{2D}{\eta_x^2} (t_j - t_i)\} \psi(x, t_j) dt_i dt_j \right] dv_x = 1.$$

We write Ex. (A2.22) similarly to Ex. (A1.22)

$$\int_{-\infty}^{\infty} \left[ \frac{1}{\sqrt{2\pi\eta_x}} \int_{-\infty}^{\infty} \psi(x, t_j) \exp\{-i \frac{v_{xj}^2}{2} \frac{2D}{\eta_x^2} t_j\} dt_j \frac{1}{\sqrt{2\pi\eta_x}} \int_{-\infty}^{\infty} \psi(x, t_i) \exp\{i \frac{v_{xi}^2}{2} \frac{2D}{\eta_x^2} t_i\} dt_i \right] dv_x = 1$$

From Ex. (A2.23), by analogy with (A1.22) – (A1.25), two Fourier transforms follow

$$\phi(k_{xj}, t_j) = \frac{1}{\sqrt{2\pi\eta_x}} \int_{-\infty}^{\infty} \psi(x, t_j) \exp\{-ik_{xj} \frac{2D}{\eta_x^2} t_j\} dt_j, \tag{A2.24}$$

$$\phi(k_{xi}, t_i) = \frac{1}{\sqrt{2\pi\eta_x}} \int_{-\infty}^{\infty} \psi(x, t_i) \exp\{ik_{xi} \frac{2D}{\eta_x^2} t_i\} dt_i. \tag{A2.25}$$

The change in the mechanical energiality of the ChWP $\varepsilon_{kx}$ due to the change in its kinetic energiality at the point $x$ at the intermediate time $t = (t_j + t_i)/2$ as $\tau = t_i - t_j \to 0$ (tending to zero) is on average equal to

$$\varepsilon_{kx} = \lim_{\tau \to 0} (k_{xj} + k_{xi})/2. \tag{A2.26}$$

Therefore, instead of integrals (A2.24) and (A2.25) for any intermediate time $t$, we can write

$$\phi(\varepsilon_{kx}, t) = \frac{1}{\sqrt{2\pi\eta_x}} \int_{-\infty}^{\infty} \psi(x,t) \, exp\{-i\frac{D}{\eta_x^2}\varepsilon_{kx}t\}dt, \tag{A2.27}$$

$$\phi^*(\varepsilon_{kx}, t) = \frac{1}{\sqrt{2\pi\eta_x}} \int_{-\infty}^{\infty} \psi(x,t) \, exp\{i\frac{D}{\eta_x^2}\varepsilon_{kx}t\}dt. \tag{A2.28}$$

According to expression (51), the globally averaged change {increase (+) or decrease (−)} of the mechanical energiality of the ChWP is equal to

$$\pm \overline{<\varepsilon_k(x,t)>} = \pm \int_{-\infty}^{\infty} \rho(\varepsilon_{kx},t)\varepsilon_{kx}d\varepsilon_{kx} = \pm \int_{-\infty}^{\infty} \varphi(\varepsilon_{kx},t)\varepsilon_{kx}\varphi^*(\varepsilon_{kx},t)d\varepsilon_{kx}, \tag{A2.29}$$

where $\varepsilon_{kx}$ is a local change in the mechanical energiality of the ChWP due to a slight change in its kinetic energiality in the direction of the *X*-axis;

$\rho(\varepsilon_{kx}, t)$ is the probability distribution function (PDF) of changes $\varepsilon_{kx}$.

In order to write the Ex, (A2.29) in a coordinate representation, we perform actions similar to (A2.1) – (A2.13), taking into account the Fourier transformations (A2.27) – (A2.28) and the validity of the expressions

$$\varepsilon_{kx}^n e^{i\frac{2D}{\eta_x^2}\varepsilon_{kx}t} = \left(-i\frac{\eta_x^2}{D}\frac{\partial}{\partial t}\right)^n e^{i\frac{2D}{\eta_x^2}\varepsilon_{kx}t}, \tag{A2.30}$$

$$\varepsilon_{kx}^n e^{-i\frac{2D}{\eta_x^2}\varepsilon_{kx}t} = \left(i\frac{\eta_x^2}{D}\frac{\partial}{\partial t}\right)^n e^{-i\frac{2D}{\eta_x^2}\varepsilon_{kx}t}.$$

As a result, we obtain a coordinate representation of the globally averaged change in the mechanical energiality of the ChWP due to the change in its averaged kinetic energiality

$$\overline{<\varepsilon_k(x,t)>} = \int_{-\infty}^{\infty} \rho(\varepsilon_{kx},t)\varepsilon_{kx}d\varepsilon_{kx} = \pm i\frac{\eta_x^2}{D}\int_{-\infty}^{+\infty} \psi(x,t)\frac{\partial \psi(x,t)}{\partial t}dx. \tag{A2.31}$$

The proof of the validity of Ex. (A2.31) completely coincides with the proof of Ex. (A2.13)*.

*The similarity of the proofs of Ex.s (A2.13) and (A2.31) corresponds to the space-time symmetry between $p_x x$ and $E_x t$ (or $v_x x$ and $\varepsilon_x t$) in the de Broglie wave function $\psi = exp\{-i(p_x x - E_x t)/\hbar\} = = exp\{-i(v_x x - \varepsilon_{kx} t)m/\hbar\} = exp\{-i(v_x x - \varepsilon_{kx} t)\eta_x\}$, which underlies classical QM.

<div align="right">Appendix 3</div>

## A3 Average quantized states of the nucleus of a biological cell

We apply the Schrödinger-Euler-Poisson equation (52)

$$-\frac{3\eta_r^2}{4}\nabla^2\psi(x,y,z) + <u(x,y,z)>\psi(x,y,z) = \varepsilon\psi(x,y,z) \quad (A3.1)$$

to describe a discrete (quantum) set of averaged states of the chaotically oscillating nucleus of a living biological cell (hereinafter, the *c*-nucleus) during the interphase period (see Figure 1*b*).

Let's assume that the mechanical energiality of the chaotically shifting *c*-nucleus is constant during the entire period of observation

$$\varepsilon_n = <k_n(x,y,z,t)> + <u_n(x,y,z,t)> = const, \quad (A3.2)$$

where

$\varepsilon_n = E_n(x,y,z,t)/m_n$ is the mechanical energiality of the *c*-nucleus with mass $m_n$;

$k_n(x,y,z,t) = T_n(x,y,z,t)/m_n$ is the kinetic energiality of the *c*-nucleus;

$u_n(x,y,z,t) = U_n(x,y,z,t)/m_n$ is the potential energiality of the *c*-nucleus.

Let's consider the case when deviations of the *k*-nucleus from the center of the cell lead to the fact that a cytoplasm surrounding it, on average, stretches so that it tends to return the *c*-nucleus to its original central position (see Figures 1*a,b*). At the same time, the elastic tensions of the cytoplasm $\sigma_v$ increase on average in proportion to the distance of the *c*-nucleus from the conditional center

$$\sigma_v(\vec{r}) \approx k_u\, r, \quad (A3.3)$$

where $k_u = K_u/m_n$ is the massless coefficient of elastic tension of the cytoplasm, $K_u$ is the force constant of the cytoplasm; $r = \sqrt{x^2 + y^2 + z^2}$ is distance from the conventional center to the *c*-nucleus (see Figures 1*a,b*).

In this case, the average potential energiality of the *c*-nucleus is approximately equal to

$$<u_n(\vec{r})> \approx \int k_u\, r\, dr = \frac{1}{2} k_u r^2. \qquad (A3.4)$$

We write equation (A3.1) taking into account the above notation in the spherical coordinate system $r, \theta, \phi$

$$\nabla^2 \psi(r,\theta,\phi) + \frac{2}{\eta_{n1}^2}\left[\varepsilon_n - \frac{k_u r^2}{2}\right]\psi(r,\theta,\phi) = 0, \qquad (A3.5)$$

where

$$\nabla^2 = \frac{1}{r^2}\frac{\partial}{\partial r}\left(r^2 \frac{\partial}{\partial r}\right) + \frac{1}{r^2}\left[\frac{1}{\sin\theta}\frac{\partial}{\partial \theta}\left(\sin\theta\frac{\partial}{\partial \theta}\right) + \frac{1}{\sin^2\theta}\frac{\partial^2}{\partial \phi^2}\right]$$

is the Laplace operator in spherical coordinates;

$$\eta_{n1} = \sqrt{\frac{3}{2}}\eta_n = \sqrt{\frac{3}{2}\frac{2\sigma_{nr}^2}{\tau_{nr}}} = \frac{\sqrt{6}\sigma_{nr}^2}{\tau_{nr}} \qquad (A3.6)$$

is the adjusted scale parameter;

$$\sigma_{nr} = \frac{1}{\sqrt{3}}\sqrt{\sigma_{nx}^2 + \sigma_{ny}^2 + \sigma_{nz}^2} \qquad (A3.7)$$

is the average standard deviation of the geometric center of the chaotically oscillating *c*-nucleus from the conditional center of the stochastic system (in the case under consideration, from the center of the biological cell, see Figure 1*b*);

$$\tau_{nr} = \frac{1}{3}(\tau_{nx} + \tau_{ny} + \tau_{nz}) \qquad (A3.8)$$

is the average autocorrelation interval of the considered three-dimensional stationary random process, i.e. random trajectory along which the geometric center of the chaotically oscillating *c*-nucleus moves.

The solutions of this equation, as is known [34], are a discrete set of wave functions depending on three quantum numbers *k*, *l*, *m*:

$$\psi_{klm}(r,\theta,\phi) = \sqrt{\sqrt{\frac{2}{\pi}\beta^3}\frac{2^{k+2l+3}k!\beta^l}{(2k+2l+1)!!}} r^l e^{-\beta r^2} L_k^{\left(l+\frac{1}{2}\right)}(2\beta r^2) Y_{lm}(\theta,\phi), \qquad (A3.9)$$

where $\beta = \frac{\sqrt{k_u}}{2\eta_{n1}} = const$ is the system parameter; $L_l^{\left(l+\frac{1}{2}\right)}(2\beta r^2)$ is generalized Laguerre polynomials; $Y_{lm}(\theta,\phi)$ is spherical harmonic function; $k = 1,2,3,...$ is the main quantum number; $l = 0,1,2,..., k–1$ is the orbital quantum number; $m = –l,...,l$ is the peripheral quantum number (in atomic physics, the number $m$ is called the "magnetic quantum number", but this name is not suitable for mass-independent stochastic quantum mechanics. Therefore, in this article, the number $m$ is proposed to be called the "peripheral quantum number"{see § 3.6 in [27]}.

Wave functions (A3.9) correspond to the eigenvalues of the mechanical energiality of the *c*-nucleus [34]

$$\varepsilon_{nkl} = \eta_{n1}\sqrt{k_u}\left(2k + l + \frac{3}{2}\right). \qquad (A3.10)$$

For example, let's estimate the main characteristics of the stochastic system under consideration (i.e., the randomly wandering geometric center of the *c*-nucleus) for $k = 1$ and $l = 0$.

Для наглядности оценим основные характеристики рассматриваемой стохастической системы (т.е. хаотически блуждающего геометрического центра к-ядра) при $k = 1$ и $l = 0$.

Let: $\sigma_{nr} \approx 1.7 \times 10^{-6} m$, $\tau_{nr} \approx 4.2 \times 10^{-3} s$, $k_u = 2.3 \times 10^{-13} s^{-2}$,

then $\qquad \eta_{n1} \approx \frac{\sqrt{6}\sigma_{nr}^2}{\tau_{nr}} \approx \frac{\sqrt{6}\times(1{,}7\times10^{-6})^2}{4{,}2\times10^{-3}} \approx 1.69 \times 10^{-9} \frac{m^2}{s}, \qquad (A3.11)$

$\varepsilon_{n10} = \eta_{n1}\sqrt{k_u}\left(N + \frac{3}{2}\right) \approx 1.69 \times 10^{-9} \times 2.3 \times 10^{-13} \times \left(2 + \frac{3}{2}\right) \approx 1.3 \times 10^{-21} \frac{m^2}{s^2}.$

As applied to the case under consideration, $|\psi_{klm}(r,\theta,\phi)|^2$ are the probability density functions (PDFs) of the possible location of the geometric center of the *c*-nucleus inside the biological cell at and different values of the quantum numbers $k$, $l$, and $m$.

To experimentally fix the averaged discrete (quantum) state of the stochastic system (i.e., a chaotically wandering $c$-nucleus), it is necessary:

- make a video recording of the chaotic behavior of the $c$-nucleus over a prolonged period of time;

- to reproduce the digitized, software-cleaned video recording of the chaotic behavior of the geometric center of the $c$-nucleus at high speed. The rate of reproduction of the motion of the geometric center of the $c$-nucleus must be so high that the given point must be "blurred" over the entire observation area.

- if all the above actions can be performed with a sufficiently high resolution and with the maximum possible exclusion of various interfering factors, then a dark-light spot configuration may appear on the monitor screen (similar to one of the spots shown in Figure 1$b$). The configuration of such a spot must correspond to the PDF of the location of the geometric center of the $c$-nucleus.

If the prediction is correct, then stochastic quantum biophysics ($\sim 10^{-3}$cm) should not differ from quantum physics of elementary particles ($\sim 10^{-13}$cm). The difference lies only in the scale of the processes under consideration, which differ from each other by about 10 orders of magnitude.

At the same time, the mass-independent stochastic quantum mechanics developed in this article predicts the possibility of similar averaged quantum states of a chaotically shifting geometric (or hypothetical) center, for example, of a planetary core, or of a galactic center, or of a bird flying in a spherical cage, or of a fluctuating stock price in 3 freely convertible currencies, etc.

It is easy, for example, to film a butterfly flying randomly around a glowing lamp or a flower bud swaying in the wind on a video camera, and then scroll the video at high speed. As a result, stable light-dark configurations are expected to be found on the screen, corresponding to a circle, or an ellipse, or a figure-of-eight or other Lissajous curve, depending on the average intensity of exposure on the butterfly or on the flower bud.

We assume that the stochastic Schrödinger-Euler-Poisson equation (52) is applicable to describe the averaged states of many such 3-dimensional chaotic processes. To do this, in each case, it is sufficient to estimate three averaged parameters of the stationary random process under study: the standard deviation $\sigma_r$, the autocorrelation interval $\tau_r$, and the massless coefficient of elastic tension of the physical (or social) medium $k_u$ surrounding the chaotically wandering object (or subject) under consideration.